\documentclass[12pt]{article}
\usepackage{amssymb,amsmath,enumerate,amsthm}
\usepackage{ushort}
\usepackage{graphicx}
\usepackage{epsfig}
\usepackage{epstopdf}
\usepackage{latexsym}
\usepackage{psfrag}
\usepackage{booktabs}
\usepackage{xspace}
\usepackage{bbm}

\usepackage[toc]{appendix}

\usepackage{color}
\usepackage{datetime}
\usepackage[
      colorlinks=false,
      linkcolor=darkblue,  
      urlcolor=blue,    
      filecolor=blue,     
      citecolor=red,
linktocpage=true,
      pdfstartview=FitV,
      bookmarksopen=true    
      ]{hyperref}

\ifpdf
\fi

\numberwithin{equation}{section}  

\def\be{\begin{equation}}
\def\ee{\end{equation}}

\def\cD{{\cal D}}

\def\cL{{\cal L}}
\def\cM{{\cal M}}

\def\cO{{\cal O}}

\def\cS{{\cal S}}

\def\cW{{\cal W}}

\def\RR{\mathbb{R}}

\def\SS{\mathbb{S}}

\definecolor{cardinal}{rgb}{0.6,0,0}
\definecolor{darkgreen}{rgb}{0,0.5,0}
\definecolor{golden}{rgb}{0.92, 0.7, 0}
\definecolor{midnight}{rgb}{0, 0, 0.5}
\definecolor{darkblue}{rgb}{0.2, 0, 0.8}
\definecolor{white}{rgb}{1,1,1}
\definecolor{black}{rgb}{0,0,0}
\definecolor{purple}{rgb}{0.4,0,0.5}

\newcommand{\dd}{\mathrm{d}}					
\newcommand{\ddh}{\hat{\mathrm{d}}}					
\DeclareMathOperator*{\hodge}{\star}				
\newcommand{\vol}{\mathrm{vol}}				
\DeclareMathOperator{\Lie}{\mathcal{L}}
\DeclareMathOperator*{\diag}{\mathrm{diag}}		
\DeclareMathOperator{\Span}{\mathrm{span}}		

\DeclareMathOperator{\ins}{\iota}					
\newcommand{\norm}[1]{{\lVert {#1} \rVert}}			

\usepackage{forloop}
\newcounter{ct}

\newcommand{\mink}[1]{(\mathord{-} \forloop[-1]{ct}{#1}{\value{ct} > 1}{\, \mathord{+}})}

\newcommand{\eucl}[1]{(%
	\ifthenelse{#1 > 0}{%
		\mathord{+} \forloop[-1]{ct}{#1}{\value{ct} > 1}{\, \mathord{+}}%
	}{}%
)}

\newcommand{\sig}[2]{(%
	\ifthenelse{#2 > 0}{%
		\mathord{-}  \forloop[-1]{ct}{#2}{\value{ct} > 1}{\, \mathord{-}}%
		\ifthenelse{#1 > 0}{\,}{}%
	}{}%
	\ifthenelse{#1 > 0}{%
		\mathord{+} \forloop[-1]{ct}{#1}{\value{ct} > 1}{\, \mathord{+}}%
	}{}%
)}

\newcommand{\sref}[1]{Section~\ref{#1}}

\newcommand{\dref}[1]{Definition~\ref{#1}}

\newcommand{\aref}[1]{Appendix~\ref{#1}}

\usepackage{empheq}
\newcommand*\padbox[1]{\fbox{\hspace{0.3em}#1\hspace{0.3em}}}
\empheqset{box=\padbox}

\theoremstyle{definition}
\newtheorem{definition}{Definition}[section]

\theoremstyle{remark}

\allowdisplaybreaks[1] 

\begin{document}  

\begin{titlepage}
 
\bigskip
\bigskip
\bigskip
\bigskip
\begin{center} 
{\Large \bf  Evanescent ergosurfaces and \\ ambipolar hyperk\"ahler metrics}

\medskip
\bigskip
\bigskip
{\bf Benjamin E.~Niehoff and Harvey S.~Reall} \\
\bigskip
DAMTP, Centre for Mathematical Sciences, \\
University of Cambridge\\
Wilberforce Road, Cambridge CB3 0WA, UK\\
\bigskip
\bigskip
{\rm B.E.Niehoff@damtp.cam.ac.uk, ~H.S.Reall@damtp.cam.ac.uk} \\

\bigskip
\bigskip 

\end{center}

\begin{abstract}

A supersymmetric solution of 5d supergravity may admit an `evanescent ergosurface': a timelike hypersurface such that the canonical Killing vector field is timelike everywhere except on this hypersurface. The hyperk\"ahler `base space' of such a solution is `ambipolar', changing signature from $\sig{4}{0}$ to $\sig{0}{4}$ across a hypersurface. In this paper, we determine how the hyperk\"ahler structure must degenerate at the hypersurface in order for the 5d solution to remain smooth. This leads us to a definition of an ambipolar hyperk\"ahler manifold which generalizes the recently-defined notion of a `folded' hyperk\"ahler manifold. We prove that such manifolds can be constructed from `initial' data prescribed on the hypersurface. We present an `initial value' construction of supersymmetric solutions of 5d supergravity, in which such solutions are determined by data prescribed on a timelike hypersurface, both for the generic case and for the case of an evanescent ergosurface.


\end{abstract}

\end{titlepage}


\tableofcontents

\section{Introduction}

There has been recent interest in the mathematical literature in `folded' hyperk\"ahler manifolds \cite{Hitchin:2015H,Biquard:2015cia}.  These are 4d manifolds which are hyperk\"ahler away from some `fold' hypersurface $\cS$ on which the hyperk\"ahler structure degenerates in a prescribed way and the metric is singular.  The `folding' action is implemented by an involution symmetry, which is a discrete isometry that exchanges one side of the fold surface with the other.  One curious feature of folded hyperk\"ahler manifolds is that the metric signature on one side of the `fold' is the usual Euclidean $\sig{4}{0}$, while on the other side becomes anti-Euclidean $\sig{0}{4}$.

This sort of feature has been a recurring theme in the physics literature in the context of the `fuzzball' or `microstate geometries' program and 5-dimensional supergravity under the guise of `ambipolar hyperk\"ahler manifolds' \cite{Giusto:2004kj,Bena:2005va,Berglund:2005vb,Bena:2007kg,Gibbons:2013tqa,Bena:2013ora,Bena:2013dka}. The working notion of an ambipolar hyperk\"ahler manifold has been ``any manifold with hyperk\"ahler structure whose metric is allowed to flip signature from $\sig{4}{0}$ to $\sig{0}{4}$ across some singular surface'', although a precise definition has thus far been lacking.  However, it has been observed that one can construct 5-dimensional supersymmetric solutions on an ambipolar hyperk\"ahler base space, where the critical surface $\cS$ is in fact \emph{not} singular from the 5d standpoint, i.e., the 5d metric is everywhere smooth with Lorentzian $\mink{5}$ signature.  This is possible because in the 5d metric, the 4d base metric is multiplied by a conformal factor which precisely cancels both the singular behavior and the change of sign.

This signature-flipping is actually quite important to the fuzzball program for the following reason. Supersymmetric solutions of 5d supergravity are constructed from a hyperk\"ahler `base space' \cite{Gauntlett:2002nw}. Hyperk\"ahler manifolds enjoy a uniqueness theorem:  the only complete hyperk\"ahler manifold asymptotic to $\RR^4$ is $\RR^4$.  A microstate geometry is a supergravity solution (in 5 or more dimensions) that has no horizons and no singularities, but which has asymptotic charges like a black hole, sourced by fluxes and non-trivial homology cycles \cite{Bena:2007kg,Gibbons:2013tqa}.  In order to have any such structure, one requires more flexibility in the base space metric than being merely $\RR^4$.  Thus \emph{asymptotically flat} microstate geometries are \emph{required} to be built on something more general than a complete hyperk\"ahler manifold.

A further 5-dimensional phenomenon associated with ambipolar base spaces is the notion of an \emph{evanescent ergosurface} \cite{Gibbons:2013tqa}, which occurs at the critical surface $\cS$.  An ordinary ergosurface is a timelike surface which is the boundary of an \emph{ergoregion}:  in an ergoregion, an asymptotically-timelike Killing vector becomes spacelike; thus the ergosurface is the transition surface on which that Killing vector is null. Supersymmetric solutions of 5d supergravity always admit a non-spacelike Killing vector field, and hence such solutions do not admit ergoregions. An \emph{evanescent} ergosurface, then, is an ergosurface \emph{without} a corresponding ergoregion:  a timelike surface such that the canonical Killing vector is timelike everywhere except on this surface, where it is null.\footnote{It is also possible for the canonical Killing vector field to be timelike everywhere except on a {\it null} hypersurface; this is the case of a supersymmetric Killing horizon, which was analyzed in Ref. \cite{Reall:2002bh}.}

The conditions under which a signature-flip of the base space is allowed have been studied only in special cases\footnote{Most of these references consider only base spaces which are a Gibbons-Hawking space \cite{Gibbons:1979zt}, although \cite{Bena:2007ju} considers more general metrics.} \cite{Bena:2007kg,Bena:2007ju,Gibbons:2013tqa,Bena:2013ora}, and have not been spelled out in general.  In this paper, we seek to remedy this situation.  We will give a precise definition of an `ambipolar hyperk\"ahler manifold' which generalizes the folded hyperk\"ahler manifolds of \cite{Hitchin:2015H,Biquard:2015cia} to the case of critical surfaces \emph{without} an involution symmetry.  We present a method for constructing such manifolds. This is based on work of Ashtekar, Jacobson and Smolin (AJS) \cite{Ashtekar:1987qx}, which provides an `inital value' construction of hyperk\"ahker manifolds from `initial data' prescribed on a hypersurface $\cS$. Biquard has shown that the same method can be used to construct a folded hyperk\"ahker manifold from data prescribed on the singular hypersurface $\cS$ \cite{Biquard:2015cia}. We will show that this method can be generalized to construct ambipolar hyperk\"ahler manifolds from the data on $\cS$. In all cases, the free data is equivalent to specifying two functions on $\cS$. 

Next we demonstrate the relevance of our definition for 5d supegravity. We focus on 5d minimal supergravity, whose bosonic sector consists of the metric $g$ and a Maxwell field $F$. We show that, if $(g,F)$ are smooth, admit a supercovariantly constant spinor, and there exists an evanescent ergosurface, then the base space must satisfy our definition of an  ambipolar hyperk\"ahler manifold. The singular surface $\cS$ corresponds to the evanescent ergosurface in 5d. In addition to the base space, the 5d solution is built from a scalar field and 1-form defined on this base space \cite{Gauntlett:2002nw} and we show how smoothness of the 5d solution determines the behaviour of these quantities near $\cS$. We show that these necessary conditions are also sufficient: given an ambipolar hyperk\"ahler space, and a 1-form and scalar with appropriate behaviour near $\cS$ one can recover 5d fields $(g,F)$ with the properties just listed.  

Usually one demands more then the existence of a supercovariantly constant spinor - one would also like to satisfy the field equations. We show that these equations do not impose any further restrictions on the base space beyond the condition that it be an ambipolar hyperk\"ahler manifold. To do this, we extend `initial value' construction of the base space to an initial value construction of a full 5d solution from data specified on $\cS$. To warm up, we show how to extend the AJS method to determine the full 5d solution from data prescribed on a non-singular hypersurface $\cS$ within a hyperk\"ahler base space, which corresponds to a timelike hypersurface in 5d. The resulting solution is specified by 8 free functions on $\cS$ (equivalent to 4 degrees of freedom in 4d). We then show how this can be extended to the ambipolar case, for which $\cS$ is singular. The resulting 5d solution is smooth with an evanescent ergosurface at $\cS$. In this case, the solution is still specified by 8 free functions on $\cS$, so the existence of an evanescent ergosurface does not impose functional constraints on a solution. 

This paper is structured as follows:  In \sref{sec:biquard}, we review the `folded hyperk\"ahler metrics' of \cite{Hitchin:2015H,Biquard:2015cia}.  In \sref{sec:ambipolar}, we give a precise definition for `ambipolar hyperk\"ahler metrics' and show how to construct them from data on $\cS$.  In \sref{sec:evanescent}, we discuss `evanescent ergosurfaces' in 5d minimal supergravity, and demonstrate the connection to ambipolar hyperk\"ahler base manifolds.  In \sref{sec:ashtekar}, we present an `initial value' construction for supersymmetric solutions of 5d supergravity, which is naturally suited to solutions in the neighborhood of an evanescent ergosurface. Finally, in \sref{sec:conclusions}, we discuss our results.

\section{Folded hyperk\"ahler metrics}
\label{sec:biquard}

\subsection{Example and definition}

In this section, we review `folded' hyperk\"ahler manifolds as defined in \cite{Hitchin:2015H,Biquard:2015cia}.  
The canonical example is a particular Gibbons-Hawking metric:
\begin{equation} \label{canonical}
h =  \frac{1}{z} \, (\dd \psi + A)^2 + z \, (\dd x^2 + \dd y^2 + \dd z^2), \qquad \dd A = \dd x \wedge \dd y.
\end{equation}
The triplet of K\"ahler 2-forms are given by
\begin{align}
X^1 &= (\dd \psi + A) \wedge \dd x - z \, \dd y \wedge \dd z, \\
X^2 &= (\dd \psi + A) \wedge \dd y - z \, \dd z \wedge \dd x, \\
X^3 &= (\dd \psi + A) \wedge \dd z - z \, \dd x \wedge \dd y.
\end{align}
We see that $h$ is undefined at $z = 0$, has signature $\sig{4}{0}$ for $z > 0$, and signature $\sig{0}{4}$ for $z < 0$. Under the involution $\iota : z \mapsto -z$, we have
\begin{equation}
\iota^* h = -h, \qquad \iota^* X^1 = X^1, \qquad \iota^* X^2 = X^2, \qquad \iota^* X^3 = - X^3.
\end{equation}
While $h$ is undefined at $z = 0$, the 2-forms $X^1, X^2, X^3$ are smooth there.  Pulling them back to 2-forms on $\cS$, we have
\begin{equation}
\cS^* X^1 = \theta \wedge \dd x, \qquad \cS^* X^2 = \theta \wedge \dd y, \qquad \cS^* X^3 = 0, \qquad \text{where} \quad \theta \equiv \dd \psi + A.
\end{equation}
Noting that $\dd \theta = \dd x \wedge \dd y$, we see that
\begin{equation} \label{contact}
\theta \wedge \dd \theta = \dd \psi \wedge \dd x \wedge \dd y \neq 0,
\end{equation}
and hence $\theta$ is a contact form on $\cS$.


From this canonical example, Hitchin \cite{Hitchin:2015H} extracts a notion of a `folded' hyperk\"ahler manifold.  A formal definition has been given by Biquard \cite{Biquard:2015cia}:

\begin{definition}[Biquard] \label{biquard def}
A \emph{folded hyperk\"ahler structure} consists of a smooth 4-manifold $\cM$, a smooth imbedded hypersurface $\cS \subset \cM$ (the fold surface), three  smooth, closed, 2-forms $X^i$ on $\cM$, and a smooth diffeomorphism $\iota: \cM \to \cM$ such that
\begin{enumerate}
\item
$\cS$ divides $\cM$ into two disjoint connected components: $\cM \setminus \cS \simeq \cM^+ \cup \cM^-$;
\item
the 2-forms $X^i$ define a hyperk\"ahler structure on $\cM^\pm$ with hyperk\"ahler metric $h^\pm$ where $h^+$ has signature $\sig{4}{0}$ and $h^-$ has signature $\sig{0}{4}$;
\item
on the surface $\cS \subset \cM$, one has
$\cS^* X^1 \neq 0$, $\cS^* X^2 \neq 0$, $\cS^* X^3 = 0$ and the distribution $\cD \subset T\cS$ given by
$\cD \equiv \ker \cS^* X^1 \oplus \ker \cS^* X^2$ is a contact distribution.\footnote{Note that $\cS^* X^1$ and $\cS^* X^2$ are non-vanishing 2-forms on the 3-manifold $\cS$, which implies that they have 1-dimensional kernels.} 
\item $\iota$ is an involution that fixes $\cS$ and maps $\cM^\pm$ to $\cM^\mp$ such that 
\begin{equation} \label{involution}
\iota^* h^\pm = - h^\mp, \qquad \iota^* X^1 = X^1, \qquad \iota^* X^2 = X^2, \qquad \iota^* X^3 = - X^3.
\end{equation}
\end{enumerate}
\end{definition}

\subsection{Construction of folded hyperk\"ahler manifolds}
\label{sec:biquard ex uniq}

In Ref. \cite{Biquard:2015cia}, Biquard gives an `initial value' construction of folded hyperk\"ahler manifolds. Given a 3-manifold $\cS$ and a pair of closed 2-forms $Y^1$ and $Y^2$ on $\cS$ such that $\cD = \ker Y^1 \oplus \ker Y^2$ is a contact distribution, he constructs, for small enough $\epsilon>0$, a folded hyperk\"ahler structure on the manifold $\cM = (-\epsilon, \epsilon) \times \cS$ such that $\cS^* X^1 = Y^1$ and $\cS^* X^2=Y^2$, where we identify $\cS$ with $\{0\} \times \cS \subset \cM$. 

In more detail, Biquard argues that one can define 1-forms $\theta,\rho^1,\rho^2$ on $\cS$ such that $\theta$ is a contact form for the distribution $\cD$ (hence $\theta \wedge d\theta \ne 0$), $d \theta = \rho^1 \wedge \rho^2$, $Y^1 = \rho^2 \wedge \theta$, $Y^2 = \theta \wedge \rho^1$. He then introduces a coordinate $x \in (-\epsilon,\epsilon)$ so that the involution acts via $\iota : x \mapsto -x$ and $\cS$ is the surface $x=0$ in $\cM$. The 2-forms that he constructs can be expanded around $\cS$ as
\begin{equation} \label{biquard omegas}
\begin{split}
X^1 &= x \, \dd x \wedge \rho^1 + \rho^2 \wedge \theta + \cO(x^2), \\
X^2 &= x \, \dd x \wedge \rho^2 + \theta \wedge \rho^1 + \cO(x^2), \\
X^3 &= \dd x \wedge \theta + x \, \rho^1 \wedge \rho^2 + \cO(x^2).
\end{split}
\end{equation}
and the metric can be expanded around $\cS$ as
\begin{equation} \label{biquard metric}
h = x^{-1} \, \theta^2 + x \Big( \dd x^2 + (\rho^1)^2 + (\rho^2)^2 \Big) + \cO(x^3)(\dd x, \rho^1, \rho^2, x^{-1} \theta),
\end{equation}
The final term denotes terms quadratic in $(\dd x, \rho^1, \rho^2, x^{-1} \theta)$ with coefficients of order $x^3$. 

Biquard's construction is a modification of the Ashtekar-Jacobson-Smolin (AJS) initial value construction of hyperk\"ahler manifolds \cite{Ashtekar:1987qx,Dancer:1992kn,Grant:1992ba}, which we review briefly here and it more detail in \aref{app:ashtekar}.  The AJS construction consists of choosing three linearly independent vector fields $V_i$ on $\cS$ which preserve a fixed volume form $v$ on $\cS$.  One then extends these vector fields off $\cS$ using the Nahm evolution equations
\begin{equation} \label{nahm eqns}
\frac{\partial}{\partial x} V_1 + [V_2, V_3] = 0, \qquad
\frac{\partial}{\partial x} V_2 + [V_3, V_1] = 0, \qquad
\frac{\partial}{\partial x} V_3 + [V_1, V_2] = 0,
\end{equation}
and, defining a fourth vector $V_0 = \partial / \partial x$, one obtains the hyperk\"ahler 2-forms and metric via
\begin{equation} \label{ashtekar formula}
 \qquad X^i = \dd x \wedge h(V_i) + i_{V_i} v, \qquad h(V_\mu, V_\nu) = v(V_1, V_2, V_3) \, \delta_{\mu\nu},
\end{equation}
Because the $V_i$ preserve $v$, one finds that the coordinate $x$ is always harmonic with respect to $h$.

To apply this method, Biquard defines $(\eta_1, \eta_2, \eta_3)$ to be the frame of vector fields on $\cS$ dual to $(\rho^1, \rho^2, \theta)$.  The fact that $Y^1$ and $Y^2$ are closed implies that $\eta_1, \eta_2$ preserve the volume form $v = \theta \wedge \dd \theta = \rho^1 \wedge \rho^2 \wedge \theta$.  One then solves Nahm's equations subject to the initial conditions
\begin{equation} \label{nahm init 1}
V_1(0) = \eta_1, \qquad V_2(0) = \eta_2, \qquad V_3(0) = 0.
\end{equation}
Standard theorems guarantee existence and uniqueness of a solution of Nahm's equations for $x \in (-\epsilon,\epsilon)$ for sufficiently small $\epsilon$. It is easy to see that $\{ V_1(-x),V_2(-x),-V_3(-x) \}$ is a solution with the same initial data and hence uniqueness implies that $V_1$, $V_2$ must be even and $V_3$ must be odd. The condition $\dd \theta = \rho^1 \wedge \rho^2$, combined with this parity symmetry gives
\begin{equation} \label{nahm init 2}
V_3(x) = x \, \eta_3 + \cO(x^3).
\end{equation}
The 2-forms \eqref{biquard omegas} and metric \eqref{biquard metric} are then obtained from the formulae \eqref{ashtekar formula}. The existence of the involution follows from the parity symmetry. The only difference from the AJS construction is that the vector fields are not linearly independent on $\cS$. This difference gives a folded hyperk\"ahler manifold instead of a  hyperk\"ahler manifold.

In summary, a folded hyperk\"ahler manifold $\cM$ can be constructed from the data on the fold surface $\cS$. It would be nice to have a proof of (local) uniquenes of this manifold (up to diffeomorphisms, extendibility etc). In other words, could there be some other folded hyperk\"ahler manifold with the same data on $\cS$? For the case of standard hyperk\"ahler manifold, the answer is no: if one defines $x$ to be a harmonic coordinate which vanishes on $\cS$ then one can recover the Nahm equations (see \aref{app:ashtekar}) and uniqueness follows from uniquess of solutions of the Nahm equations. The same would be true in the folded hyperk\"ahler case if one could argue that it is possible to choose a harmonic coordinate that vanishes on $\cS$ \cite{Biquard:2015cia}. However, proving this looks non-trivial because the harmonic condition depends on the metric, which is singular at $\cS$. 


%

\section{`Ambipolar' hyperk\"ahler metrics}
\label{sec:ambipolar}

\subsection{Motivation and definition}

The definition of a folded hyperk\"ahler manifold was motivated by the example \eqref{canonical}. Consider now a general Gibbons-Hawking metric: 
\begin{equation} \label{GH ansatz}
h = \frac{1}{V} (\dd \psi + A)^2 + V \, (\dd x^2 + \dd y^2 + \dd z^2), \qquad \dd A = \hodge_3 \dd V,
\end{equation}
If $V$ vanishes on some surface $\cS$ then the metric near $\cS$ has some similarity with \eqref{biquard metric}. However, in general this will not satisfy the definition of a folded hyperk\"ahler manifold because it lacks the involution symmetry \eqref{involution}. For example, one could consider a case for which $V = 0$ on a sphere in $\RR^3$, such as the negative-mass Taub-NUT metric, with
\begin{equation}
V = 1 - \frac{m}{r}, \qquad r \equiv \sqrt{x^2 + y^2 + z^2}.
\end{equation}
In this example, the hyperk\"ahler 2-forms are smooth at the surface $r=m$ which partitions the manifold into regions $\cM^+, \cM^-$ in which the metric has $\sig{4}{0}$ or $\sig{0}{4}$ signature respectively. Given that such manifolds play an important role in 5d supergravity, it is desirable to generalize definition \ref{biquard def} to encompass such examples. We will adopt the following definition:

\begin{definition} \label{ambipolar def}
An \emph{ambipolar hyperk\"ahler structure} consists of a smooth 4-manifold $\cM$, a smooth imbedded hypersurface $\cS \subset \cM$, and three  smooth, closed, 2-forms $X^i$ on $\cM$, such that
\begin{enumerate}
\item
$\cS$ divides $\cM$ into two disjoint connected components: $\cM \setminus \cS \simeq \cM^+ \cup \cM^-$;
\item
the 2-forms $X^i$ define a hyperk\"ahler structure on $\cM^\pm$ with hyperk\"ahler metric $h^\pm$ where $h^+$ has signature $\sig{4}{0}$ and $h^-$ has signature $\sig{0}{4}$;
\item
(a) At each point of $\cS$, the subspace $\cW =\Span \{\cS^* X^1, \cS^* X^2, \cS^* X^3\}$ of $\Lambda^2 T^* \cS$ is 2-dimensional. (b) Let $\cD$ be the union of the kernels of the non-zero elements of $\cW$. Then $\cD$ is a contact distribution.
\end{enumerate}
\end{definition}
Point 3(b) may need a little more explanation. Introduce a basis $\{\beta^1, \beta^2\}$ for $\cW$. The 2-forms $\beta^1$, $\beta^2$ are non-zero and therefore have 1-dimensional kernels (as $\cS$ is 3-dimensional). Let the vectors $\eta_1$, $\eta_2$ be non-zero elements of these kernels. Choose another vector $\eta_3$ such that $\{\eta_1,\eta_2,\eta_3\}$ is a basis for the tangent space of $\cS$. Let $\{\theta^i\}$ denote the dual basis of 1-forms. Then $\beta^1$ is proportional to $\theta^2 \wedge \theta^3$ and $\beta^2$ is proportional to $\theta^1 \wedge \theta^3$ so $\cW$ is the set of 2-forms of the form $(a_1 \theta^1 + a_2 \theta^2) \wedge \theta^3$. It is then easy to see that $\cD = \Span \{\eta_1,\eta_2\}$, so, at any point, $\cD$ is a 2-dimensional subspace of the tangent space of $\cS$. The non-trivial content of point 3(b) of our definition is that $\cD$ must be a contact distribution, i.e., $[\eta_1,\eta_2] \notin \cD$. Equivalently, $\theta_3$ must be a contact form, i.e.,  
\begin{equation} \label{ambi struct 1}
\theta^3 \wedge \dd \theta^3 \neq 0.
\end{equation}
 
Compared to the definition of a folded hyperk\"ahler structure, we have eliminated condition 4 and weakened condition 3. Compared to previous work in the supergravity literature, we have, in point 3, specified precisely {\it how} the hyperk\"ahler structure should degenerate on $\cS$. 
 
\subsection{Construction of ambipolar hyperk\"ahler manifolds}

\label{construction}

We will now show how to construct an ambipolar hyperk\"ahler manifold given the data on $\cS$. The method is a generalization of Biquard's construction of folded hyperk\"ahler manifolds. 

Let $\cS$ be an oriented 3-manifold and let $Y^i$, $i=1,2,3$, be closed 2-forms on $\cS$ such that $\cW \equiv \Span\{Y^1,Y^2,Y^3 \}$ is everywhere 2-dimensional. Let $\cD$ be the union of the kernels of the non-zero elements of $\cW$. Assume that $\cD$ is a contact distribution. We will construct, for small enough $\epsilon>0$, an ambipolar hyperk\"ahler structure on the manifold $\cM = (-\epsilon, \epsilon) \times \cS$ such that $\cS^* X^i = Y^i$, where we identify $\cS$ with $\{0\} \times \cS \subset \cM$. 

If $Y^i \ne 0$ then it has a 1-dimensional kernel inside $\cD$; let the vector field $t_i$ on $\cS$ be a non-zero element of this kernel. If $Y^i=0$ then we define $t_i=0$. The vector fields $t_i$ are linearly dependent and span $\cD$. Now pick an arbitrary volume form $v$ on $\cS$. If $t_i$ is non-zero then the 2-form $\ins_{t_i} v$ has a 1-dimensional kernel containing $t_i$. This implies that it is a multiple of $Y^i$. Obviously the same holds if $t_i=0$. Hence by rescaling $t_i$ appropriately we can arrange that
\be
\label{Videf}
 \ins_{t_i} v = Y^i
\ee
which implies that the $t_i$ are divergence-free w.r.t. $v$:
\be
 \Lie_{t_i} v = \dd \left( \ins_{t_i} v \right) = \dd Y^i = 0.
\ee
The idea now is to define vector fields $V_i$ on $\cM$ by solving Nahm's equations \eqref{nahm eqns} with initial data
\be
\label{initial}
 V_i|_{x=0} =t_i.
\ee
We then define $V_0 =\partial/\partial x$. The volume form $v$ is extended into $\cM$ by Lie transport w.r.t. $V_0$. The metric and 2-forms $X^i$ given by \eqref{ashtekar formula} will then satisfy our definition of an ambipolar hyperk\"ahler structure. We will now show this in more detail. 

Let $\{\eta_1,\eta_2\}$ be a basis for $\cD$. We can expand our vector fields $t_i$ in terms of this basis:\footnote{Latin indices $a,b,c, \ldots$ will take the values $1,2$.}
\be
 t_i = t_i^a \eta_a
\ee
Since $t_i^a t_i^b$ is positive-definite, we can normalize the basis vectors $\eta_a$ so that
\be
\label{tnorm}
 t_i^a t_i^b = \delta^{ab}
\ee
Since the $Y^i$ are linearly dependent there exists a map $u : \cS \to \SS^2$ that tells us which linear combination of them vanishes:
\begin{equation} \label{ambi struct 2}
u_i Y^i = 0, \qquad u_i u_i = 1
\end{equation}
Equation \eqref{Videf} implies that $u_i t_i = 0$ and hence
\be
 u_i t_i^a=0
\ee
We can regard $t^1_i$, $t^2_i$ and $u_i$ as orthonormal vectors in $\mathbb{R}^3$. The overall sign of $u_i$ is arbitrary; we fix this sign by demanding
\be
\label{usign}
 \epsilon_{ijk} t_i^a t_j^b u_k = \epsilon^{ab}
\ee
We now extend $\eta_a$ to a basis $\{\eta_1,\eta_2,\eta_3\}$ of vector fields on $\cS$. There is freedom in choosing $\eta_3$: we could just as well use
\be
  \eta_3' = \alpha^a \eta_a + \beta \eta_3
\ee
where $\beta \ne 0$. The condition that $\cD$ is a contact distribution is equivalent to $[\eta_1, \eta_2]^3 \neq 0$. By appropriate choices of $\alpha^a$ and $\beta$ we can arrange that $[\eta_a,\eta_3]^3=0$ and $[\eta_1, \eta_2]^3 = -1$, so we can write 
\begin{equation} \label{eta alg} 
[\eta_1, \eta_2] = \varepsilon^{ab} \kappa_b \, \eta_a - \eta_3, \qquad [\eta_a, \eta_3] = - \lambda^b{}_a \, \eta_b
\end{equation}
for certain functions $\kappa_b$ and $\lambda^b{}_a $ on $\cS$.  Since the $t_i = t_i^a \, \eta_a$ are divergence-free, it follows that the $\kappa_a$ can be written
\begin{equation} \label{kappas}
\kappa_a = t_i^b \, \eta_b(t_i^a).
\end{equation}
The precise form of the $\lambda^a{}_b$, however, will be unimportant.

Let $\{\theta^1,\theta^2,\theta^3\}$ be the dual basis, so $\theta^3$ is a contact form. In terms of the dual basis we have
\begin{equation} \label{theta alg}
\dd \theta^a = - \varepsilon^{ab} \kappa_b \, \theta^1 \wedge \theta^2 + \lambda^a{}_b \, \theta^b \wedge \theta^3, \qquad \dd \theta^3 = \theta^1 \wedge \theta^2.
\end{equation}
Next we exploit the freedom to choose the volume form $v$. If we make some other choice $v'$ then we have $v'=\lambda v$ for some non-zero function $\lambda$. This gives $t'_i = \lambda^{-1} t_i$ and hence $\eta_a' = \lambda^{-1} \eta_a$. We then find $\eta_3' = \lambda^{-2} \eta_3$. Hence $\theta^{a'} = \lambda \theta^a$, $\theta^{3'} = \lambda^2 \theta^3$ so 
\be
 v' = \lambda v = \lambda v_{123} \theta^1 \wedge \theta^2 \wedge \theta^3 = \lambda^{-3} v_{123} \theta^{1' }\wedge \theta^{2' }\wedge \theta^{3'} 
\ee
and we now choose $\lambda^3 = v_{123}$. This shows that it is consistent with our above choice of basis to pick
\be
 v = \theta^1 \wedge \theta^2 \wedge \theta^3
\ee
From equation \eqref{Videf} we now have
\be
\label{Yibasis}
 Y^i = \epsilon^{ab} t_i^a \; \theta^b \wedge \theta^3
\ee
We can now solve Nahm's equations \eqref{nahm eqns}. Writing
\be
 V_i = V_i^a \eta_a + V_i^3 \eta_3
\ee
these are a system of ODEs for $V_i^a$ and $V_i^3$. By standard theorems, there exists $\epsilon>0$ such that for $x \in (-\epsilon,\epsilon)$ there exists a unique solution of Nahm's equations obeying the initial condition \eqref{initial}, i.e.,
\be
V_i^a|_{x=0} = t_i^a \qquad V_i^3|_{x=0} = 0.
\ee
By explicit calculation we find that this solution can be expanded as
\begin{align}
V_i{}^a &= t_i^a + x \, t_i^b \, \varepsilon^{ac} \varepsilon_{bd} \, \mu_{cd} + \cO(x^2), \label{Via exp} \\
V_i{}^3 &= x (1+x \mu^a{}_a) \, u_i  + \cO(x^3), \label{Vi3 exp}
\end{align}
where the quantity
\begin{equation}
\mu_{ab} \equiv - \varepsilon_{ac} t_i^c \, \eta_b(u_i), \qquad \mu^a{}_a \equiv \delta^{ab} \mu_{ab}
\end{equation}
will appear in several places in the expansions of $h$ and $X^i$.  Note that our assumption that $\cD$ is a contact distribution ensures that $V_i{}^3$ becomes non-zero at order $x$.  This ensures that the vector fields $V_i$ are linearly independent for $x \ne 0$. 

To assemble the metric tensor, we first define
\begin{equation}
\varphi = v(V_1, V_2, V_3) = (\theta^1 \wedge \theta^2 \wedge \theta^3)(V_1, V_2, V_3),
\end{equation}
and a calculation gives
\begin{equation}
\varphi = x \big( 1 + 2 x \mu^a{}_a \big) + \cO(x^3).
\end{equation}
We now choose coordinates $y^i$ on $\cS$ and use $(x,y^i)$ as coordinates on $\cM=(-\epsilon,\epsilon) \times \cS$ and identify $\cS$ with the surface $x=0$ in $\cM$. We can regard the $V_i$ as vector fields on $\cM$ which are tangent to the level sets of $x$. We define a fourth vector field $V_0 = \partial/\partial x$. The metric and 2-forms $X^i$ on $\cM$ are then defined by \eqref{ashtekar formula}. 

More explicitly, the metric can be written as 
\be
\label{h defa}
 h = \varphi \left(dx^2 + H_{ab} \theta^a \otimes \theta^b \right) + h_{33}  \left( \theta^3 - H_{ab} V_i^3 V_i^b \; \theta^a \right)^2 
\ee
 where $H_{ab}$ is the inverse of
\be
 H^{ab} \equiv V_i^a V_i^b
\ee 
and
\be
h_{33} = \frac{\varphi}{V_i^3 V_i^3 - H_{ab} V_i^3 V_i^a V_j^3 V_j^b} 
\ee
We emphasize that the $x$-dependence of the metric arises entirely from the $x$-dependence of the $V_i$, in particular $\theta^a$ and $\theta^3$ are independent of $x$. For $x \ne 0$, the coordinate $x$ is harmonic w.r.t. $h$. 

From the expressions \eqref{Via exp}, \eqref{Vi3 exp} we see that $V_i{}^3 V_i{}^a = \cO(x^3)$, which greatly simplifies the expansions of the metric components.  Expanding in $x$, one has
\begin{equation}
H^{ab} = \delta^{ab} + 2x \, \varepsilon^{ac} \varepsilon^{bd} \mu_{cd} + \cO(x^2),
\end{equation}
and hence, since $\mu_{ab}$ is a $2\times2$ symmetric matrix,\footnote{%
To show that $\mu_{ab}$ is symmetric, use $u_i Y^i=0$, and write
$0 = d(u_i Y^i) = -d(i_{u_i t_i} v)  = -d(i_{t_i} (u_i v)) = -\cL_{t_i}(u_i v) = -\cL_{t_i}(u_i) v$ hence $\varepsilon^{ab} \mu_{ab} = - t_i^a \, \eta_a (u_i) = -\Lie_{t_i} (u_i) v =0$.
}
one obtains
\begin{equation}
\label{hexp}
H_{ab} = (1 - 2x \mu^c{}_c) (\delta_{ab} + 2x \mu_{ab}) + \cO(x^2), \qquad h_{33} = x^{-1} + \cO(x).
\end{equation}
This implies that the metric $h$ can be expanded around $x=0$ as
\begin{equation} \label{h def}
\begin{split}
h &= x^{-1} ( \theta^3 )^2 + x \big( 1 + 2 x \mu^a{}_a \big) \, \dd x^2 + x \delta_{ab} \big( \theta^a + x \mu_{ac} \, \theta^c \big) \otimes \big( \theta^b + x \mu_{bd} \, \theta^d \big) \\
& \qquad + \cO(x^3)(\dd x, \theta^a, x^{-1} \, \theta^3),
\end{split}
\end{equation}
and expanding the 2-forms $X^i$ gives
\begin{equation} \label{X def}
\begin{split}
X^i &= \dd x \wedge \Big[ (1+x \mu^c{}_c) \big( u_i \, \theta^3 + x t_i^a \, \theta^a \big) + x^2 t_i^a \mu_{ab} \, \theta^b \Big] + x u_i (1+x \mu^c{}_c) \, \theta^1 \wedge \theta^2 \\
& \qquad + \Big[(1+x \mu^c{}_c) t_i^b - x t_i^a \mu_{ab} \Big] \varepsilon^{bd} \, \theta^d \wedge \theta^3 + \cO(x^3)(\dd x, \theta^a, x^{-1} \, \theta^3).
\end{split}
\end{equation}
We now can now check that the above construction satisfies our definition. We identify the regions $\cM^\pm$ as the regions $x>0$ and $x<0$ respectively and $\cS$ as the surface $x=0$. We see that the 2-forms are smooth at $x=0$, as required. The metric has signature $\sig{4}{0}$ in $\cM^+$ and $\sig{0}{4}$ in $\cM^-$. Condition 2 of our definition is satisfied in $\cM^\pm$ because our construction reduces to the standard AJS construction of a hyperk\"ahler manifold in these regions. Finally, if we use \eqref{X def} to calculate the pullback $X^i$ to $x=0$ it agrees with our expression \eqref{Yibasis} for $Y^i$ and hence condition 3 of our definition is satisfied because of the assumed properties of the $Y^i$. 

We have construced an ambipolar hyperk\"ahler manifold given the data on $\cS$. One can now ask about (local) uniqueness of this manifold: could there be some other ambipolar hyperk\"ahler space with the same data on $\cS$? Just as for a folded hyperk\"ahler space, uniqueness would follow if one could argue that it is always possible to introduce a harmonic coordinate $x$ that vanishes on $\cS$ because one could then define vector fields $V_i$ as in \aref{app:ashtekar}, recover Nahm's equations and deduce uniqueness from uniqueness of solutions of Nahm's equations. However, as in the folded case, proving that one can define such a coordinate $x$ is non-trivial because the harmonic condition depends on the metric, which is singular on $\cS$. 

In \aref{app:ashtekar} we explain that the initial data for the standard AJS construction is equivalent to specifying $2$ functions on the initial surface $\cS$. It is interesting to see how this counting works for an ambipolar hyperk\"ahler space. Fix a coordinate chart $y^i$ on $\cS$. The solution is determined once we have chosen the vector fields $t_i$ and the volume form $v$ on $\cS$. The vector fields $t_i$ must span a 2d space, which is a single functional constraint on them. They must also be divergence free w.r.t. $v$, which is $3$ constraints. So choosing the components of the $t_i$ involves $3\times 3 -1 -3 = 5$ free functions of $y^i$. Of course there is freedom to perform coordinate transformations of the $y^i$, i.e., 3 free functions are gauge. This leaves $5-3=2$ gauge-invariant free functions. As in the standard AJS case, the freedom to choose $v$ is equivalent to a freedom in specifying the coordinate $x$, i.e., it is gauge. Hence an ambipolar hyperk\"ahler space is determined by 2 gauge invariant free functions on $\cS$, equivalent to a single "degree of freedom", exactly as for a regular hyperk\"ahler space. 

\section{Evanescent ergosurfaces in 5d supergravity}
\label{sec:evanescent}

One idea that has consistently appeared in the microstate geometry program, but was only recently given a name, is the notion of an \emph{evanescent ergosurface} \cite{Gibbons:2013tqa}.  Supersymmetric solutions of 5d supergravity admit a Killing vector field that is everywhere timelike or null \cite{Gibbons:1993xt}. An evanescent ergosurface is a timelike hypersurface such that this canonical Killing vector field is timelike outside the hypersurface but null on the hypersurface. Since the Killing vector field cannot be spacelike, there is no actual ergoregion; an evanescent ergosurface is essentially the limit of an ergoregion as it flattens out into a surface of zero thickness.

For supersymmetric microstates geometries in 5d supergravity, the existence of evanescent ergosurfaces is actually \emph{necessary} due to the uniqueness of $\RR^4$ as a strict hyperk\"ahler manifold.  The presence of such surfaces has proven to have interesting physical consequences \cite{Bena:2013ora}.  Here, however, we will show they have mathematical consequences:  the presence of an evanescent ergosurface naturally corresponds to a `base space' geometry which is an ambipolar hyperk\"ahler metric, satisfying our definition \dref{ambipolar def}.

We start by reviewing the canonical form of supersymmetric configurations of 5d minimal supergravity, as determined in \cite{Gauntlett:2002nw}. We then assume that we have a 5d supersymmetric spacetime with an evanescent ergosurface and  prove that the corresponding base space must be an ambipolar hyperk\"ahler manifold, with the $1$-form defined on this base behaving in a certain (singular) way near $\cS$. Finally, we prove the converse: given such a base space and $1$-form one obtains smooth 5d fields with an evanescent ergosurface. 

\subsection{Supersymmetric configurations of 5d minimal supergravity}

In this section we will review properties of supersymmetric configurations of 5d minimal supergravity, as determined in Ref. \cite{Gauntlett:2002nw}. We say "configurations" rather than "solutions" because many of the results of Ref. \cite{Gauntlett:2002nw} rely only on the existence of a supercovariantly constant spinor, rather than the full field equations. 

The bosonic sector of 5d minimal supergravity consists of a metric tensor $g$ and a Maxwell field $F$, with action 
\begin{equation}
S = \frac{1}{4 \pi G} \int \bigg( \frac14 \hodge_5 R - \frac12 F \wedge \hodge_5 F - \frac{2}{3 \sqrt3} F \wedge F \wedge A \bigg), \qquad F \equiv \dd A.
\end{equation}
A canonical form for supersymmetric bosonic configurations $(g,F)$ of this theory was determined in  Ref. \cite{Gauntlett:2002nw}. By definition, such a configuration admits a globally defined supercovariantly constant spinor field $\epsilon$. From $\epsilon$ one can construct a scalar field $f$, a vector field $K$ and three 2-forms $X^i$, all quadratic in $\epsilon$, satisfying the algebraic relations
\begin{align}
K_\alpha K^\alpha &= - f^2, \label{alg V} \\
\ins_K X^i &= 0, \label{alg VX} \\
\ins_K \hodge_5 X^i &= - f\, X^i, \label{alg V*X} \\
X^i \wedge X^j &= 2 \, \delta^{ij} f \hodge_5 K, \label{alg XwX} \\
X^i_{\gamma \alpha} \, X^j{}^\gamma{}_\beta &= \delta^{ij} \Big( f^2 \, \eta_{\alpha \beta} + K_\alpha K_\beta \Big) - f \, \varepsilon_{ijk} \, X^k_{\alpha \beta}, \label{alg XX}
\end{align}
where $\eta_{\alpha \beta} = \diag (-1, 1, 1, 1, 1)$ and $\hodge_5$ denotes the 5d Hodge dual. Since $f$ is real, $K$ must be timelike or null, but never spacelike.\footnote{It can be shown that $K$ cannot vanish \cite{Gauntlett:2002nw}.} 

From the Killing spinor equation, one obtains differential constraints \cite{Gauntlett:2002nw}. First, $K$ is Killing and generates a symmetry of the metric and Maxwell fields
\begin{equation}
\Lie_K g = 0, \qquad \Lie_K F = 0,
\end{equation}
where $g$ is the 5d metric. We also have
\begin{align}
\dd f &= - \frac{2}{\sqrt3} \ins_K F, \label{diff df} \\
\dd K &= - \frac{4}{\sqrt3} f \, F - \frac{2}{\sqrt3} \hodge_5 \, (F \wedge K), \label{diff dV} \\
\dd X^i &= 0, \label{diff dX} \\
\dd \hodge_5 X^i &= - \frac{2}{\sqrt3} \, F \wedge X^i. \label{diff d*X}
\end{align}
It is then easy to see that $K$ also generates a symmetry of $f$ and $X^i$. 

If $f^2>0$, then $K$ is timelike so we can introduce coordinates $(t,x^m)$ so that
\begin{equation}
K = \frac{\partial}{\partial t},
\end{equation}
and since $K$ generates a symmetry, every quantity is independent of the coordinate $t$. By taking a quotient of the 5d spacetime w.r.t. this symmetry one obtains a 4d manifold with coordinates $x^m$, referred to as the `base space'. The 5d metric can be written
\begin{equation} \label{5d metric}
g = - f^2 \, (\dd t + \omega)^2 + f^{-1} \, h,
\end{equation}
where $h = h_{mn} \, \dd x^m \dd x^n$ is a Riemannian metric on the base space and $\omega = \omega_m \, \dd x^m$ is a 1-form living on $h$.  The reason for the factor of $f^{-1}$ in front of $h$ is because then $f$ drops out of equations \eqref{alg VX}, \eqref{alg XwX} and \eqref{alg XX}:
\begin{gather}
\label{4dalg}
X^i = - \hodge_4 X^i, \qquad X^i \wedge X^j = -2 \, \delta^{ij} \, \vol_h, \\
(X^i)_m{}^p (X^j)_p{}^n = - \delta^{ij} \delta_m{}^n + \varepsilon_{ijk} \, (X^k)_m{}^n.
\end{gather}
where $\hodge_4$ is the Hodge dual w.r.t. $h$. Hence the $X^i$ define a hyperk\"ahler structure on the base space, with associated metric $h$. Note that if $f>0$ then $h$ has signature $\sig{4}{0}$ and if $f<0$ then $h$ has signature $\sig{0}{4}$. 

Equations \eqref{diff df} and \eqref{diff dV} determine the form of $F$:
\begin{equation} \label{F def}
F = \frac{\sqrt3}{2} \bigg[\dd \left(  f (\dd t +  \omega) \right)  - \frac23 \, G^+ \bigg],
\end{equation}
where 
\be
G^\pm = \frac12f  \, (1 \pm \hodge_4)  \dd \omega
\ee
So in terms of quantities which appear in the metric, 
\begin{equation} \label{F omega}
F = \frac{\sqrt3}{2} \bigg[ - (\dd t + \omega) \wedge \dd f + \frac23 f \, \dd \omega - \frac13 \, \hodge_4 f \, \dd \omega \bigg],
\end{equation}
which will be useful later.  

So far, we have assumed only the existence of a supercovariantly constant spinor for which $f \ne 0$. For the fields $(g,F)$ to be a {\it solution} of the field equations we also need to impose the equations of motion for the Maxwell field (the Einstein equation is then satisfied automatically \cite{Gauntlett:2002nw}). Together with the definition of $G^+$, this gives the `BPS equations':
\begin{align}
\dd G^+ &= 0, \label{Gplus} \\
\dd \hodge_4 \dd f^{-1} &= \frac49 \, G^+ \wedge G^+, \label{f eqn} \\
\dd \omega &= f^{-1} \, G^+ + f^{-1} \, G^-,  \label{omega eqn}
\end{align}
These equations form an upper-triangular linear system which can be solved as follows \cite{Bena:2004de} (see also \cite{Bena:2007kg}).  First one chooses a base space metric. Next one finds self-dual $G^+$ satisfying \eqref{Gplus}, and then $f$ satisfying \eqref{f eqn}. Finally one can take the exterior derivative of \eqref{omega eqn} to obtain an equation for anti-self-dual $G^-$. Solving this, one substitutes the result back into \eqref{omega eqn} to obtain an equation which can be solved for $\omega$. The metric and Maxwell field are then fully determined.

\subsection{Evanescent ergosurfaces}

We now assume that our 5d spacetime has an evanescent ergosurface. By this, we mean that there exists a smooth timelike hypersurface $\cS$ such $f(p)=0$ if, and only if, $p \in \cS$. In other words, $K$ is timelike off $\cS$ and null on $\cS$. We assume that our supercovariantly constant spinor is smooth at $\cS$, which implies that $f$, $K$ and $X^i$ are also smooth at $\cS$. We assume also that the 5d metric and Maxwell field are smooth at $\cS$. 

Since $K$ generates a symmetry, we can still take a quotient of our 5d spacetime to obtain a 4d base space. The evanescent ergosurface corresponds to a certain hypersurface in the base space which we will also call $\cS$. Away from $\cS$, we will have the structure described above, in particular the base space is hyperk\"ahler. The 2-forms $X^i$ are smooth at $\cS$ but the the hyperk\"ahler structure degenerates on $\cS$. We will show that it degenerates in precise agreement with our definition of an ambipolar hyperk\"ahler manifold. That is, our definition of an ambipolar hyperk\"ahler manifold can be thought of as naturally arising from the Killing spinor conditions of 5d minimal supergravity.

The proof is in two parts.  First, we show that smoothness of the 5d metric implies that, on $\cS$, $\cW=\Span\{\cS^* X^1, \cS^* X^2, \cS^* X^3\}$ is two-dimensional, in agreement with condition 3(a) of our definition.  Second, we show that smoothness of the Maxwell field implies that condition 3(b) is also satisfied. To do this we need one technical assumption, namely that $f$ has a {\it first order} zero on $\cS$.  This assumption can be justified by appeal to genericity; alternatively, one can use the equation of motion for the Maxwell field to show that $f$ cannot have a higher-order zero on $\cS$ (see \aref{app:general}). This is the only place where we use the equations of motion.

\subsubsection{Smoothness of the 5d metric}

We start by introducing a coordinate chart in a neighbourhood of $\cS$. We assume that the 5d spacetime is globally hyperbolic with Cauchy surface $\Sigma$. Hence each orbit of $K$ intersects $\Sigma$ exactly once. Let $T$ be the parameter distance from $\Sigma$ along orbits of $K$. We can introduce coordinates $x^m$ on $\Sigma$ and `carry' them along the integral curves of $K$ to define spacetime coordinates $(T,x^m)$. The 5d metric can be written in ADM form
\be
 \dd s^2 = -(f^2 + g_{mn} \beta^m \beta^n ) \, \dd T^2 + g_{mn}(\dd x^m + \beta^m \, \dd T) (\dd x^n + \beta^n \, \dd T)
\ee
with $K=\partial/\partial T$. Since $f$ is constant along orbits of $K$, it follows that these orbits must be tangent to $\cS$ because $\cS$ is the set of points with $f=0$.\footnote{In fact \eqref{diff dV} implies that $K \cdot \nabla K=0$ on $\cS$ so, on $\cS$, the orbits of $K$ are affinely parameterized null geodesics.} The intersection $\cS \cap \Sigma$ is a hypersurface within $\Sigma$.

Next we define a function $x$ as follows. We require that $x$ be a $K$-invariant solution of the wave equation:
\begin{equation}
\label{xeq}
\dd \hodge_5 \dd x = 0, \qquad \Lie_K x = 0,
\end{equation}
Working in the coordinates $(T,x^m)$ one sees that this is equivalent to $x$ satisfying a certain elliptic equation on $\Sigma$. By the Cauchy-Kowalevski theorem there exists, in a neighbourhood of $\cS\cap \Sigma$, a solution satisfying $x=0$ on $\cS\cap \Sigma$ and $\hat{n} \cdot \nabla x |_{\cS\cap \Sigma}= \hat{\alpha}$ where $\hat{n}$ is a unit normal to $\cS \cap \Sigma$ (w.r.t. the induced metric on $\Sigma$) and $\hat{\alpha}$ is a non-zero free function on $\cS$. From a 5d perspective, this means that we can define a function $x$ obeying \eqref{xeq} and satisfying 
\be
\label{xconds}
 x|_{\cS}=0 \qquad n \cdot \nabla x|_{\cS} = \alpha
\ee
where $n$ is a unit normal to $\cS$ (w.r.t. the 5d metric) and $\alpha$ is a non-zero $K$-invariant free function on $\cS$. 

The next step is to introduce coordinates $y^i$ on the 3d manifold $\cS \cap \Sigma$. On $\cS$ we then define $t$ to be the parameter distance from $\cS \cap \Sigma$ along the integral curves of $K$. This defines coordinates $(t,y^i)$ on $\cS$ such that $K=\partial/\partial t$ on $\cS$. Finally we extend these coordinates off $\cS$ by defining them to be constant along the integral curves of $M^\mu=g^{\mu\nu}(dx)_\nu$. This defines a coordinate chart $(t,x,y^i)$ such that $0 =M^\mu \partial_\mu t= g^{\mu\nu} (dx)_\nu \partial_\mu t = g^{tx}$ and similarly $0=g^{ix}$. This implies $0=g_{tx}=g_{ix}$. Since $K$ is a Killing field and $\Lie_K x = 0$ we have $\Lie_K M = 0$. This implies that $K = \partial/\partial t$ everywhere.

In summary, we have shown that we can introduce coordinates $(t,x,y^i)$ in a neighbourhood of $\cS$ such that $K=\partial/\partial t$ and the 5d metric takes the form
\begin{equation} \label{5d metric harm}
g = - f(x,y)^2 \, \dd t^2 - 2 \nu_i(x,y) \, \dd t \, \dd y^i + N(x,y)^2 \, \dd x^2 + \gamma_{ij}(x,y) \, \dd y^i \, \dd y^j,
\end{equation}
where $\cS$ is the surface $x=0$ hence $f = 0$ on $x=0$. All components are smooth at $x=0$. On $\cS$ we have
\be
 K|_{\cS} = -\nu_i \, \dd y^i |_{\cS}
\ee
so $\nu_i \, \dd y^i$ cannot vanish on $\cS$. $\gamma_{ij}$ is a Riemannian metric on the 3-manifold $\Sigma \cap \cS$. 
The non-zero function $N(x,y)$ is constrained by the wave equation in \eqref{xeq} which reduces to
\begin{equation}
\label{N0}
\frac{\partial}{\partial x} \Big( N^{-1} \, \sqrt{\det g_4} \Big) = 0, \qquad \text{and hence} \qquad N(x,y) = N_0(y) \sqrt{\det g_4},
\end{equation}
where $g_4$ is what remains after erasing the $\dd x^2$ term from the 5d metric \eqref{5d metric harm}.  The non-zero function $N_0(y)$ is pure gauge, and corresponds to the freedom to choose the function $\alpha(y)$ in \eqref{xconds}. 

We now define the base space manifold $\cM$ as the space of orbits of $K$ \cite{Geroch:1970nt}. We define a map $\psi: \cM_5 \rightarrow \cM$ (where $\cM_5$ is the spacetime manifold) which maps a point $p \in M_5$ to the orbit of $K$ through $p$. Since this orbit is labelled by $(x,y^i)$ we can regard $(x,y^i)$ as coordinates on $\cM$.\footnote{As an abstract manifold, $\cM$ is diffeomorphic to $\Sigma$ but we do not want to regard $\cM$ as a particular hypersurface in spacetime.} The image of $\cS$ under $\psi$ is a hypersurface in $\cM$ which we will also denote as $\cS$. Since $f$ is preserved by $K$, it can be regarded as a function on $\cM$, which is smooth everywhere, including at $\cS$. From \eqref{alg VX} and $\Lie_K X^i=0$ it follows that the 2-forms $X^i$ can also be regarded as (closed) 2-forms on $\cM$ \cite{Geroch:1970nt}. Since these 2-forms are smooth in 5d they will also be smooth on $\cS$ within $\cM$. 

If $x>0$ or $x<0$ then $f \ne 0$ so the 5d metric can also be written in the canonical form \eqref{5d metric}. The base space appearing in \eqref{5d metric} is simply the region $x>0$ or $x<0$ of $\cM$. We refer to these two regions as $\cM^\pm$. Clearly $\cM \backslash \cS \simeq \cM^+ \cup \cM^-$. From \eqref{5d metric}, we can identify the angular momentum 1-form $\omega$ and the 4d hyperk\"ahler base metric $h$ on $\cM^\pm$:
\begin{equation}
\omega \equiv f^{-2} \, \nu_i \, \dd y^i, \qquad 
h = f^{-1} \, (\nu_i \, \dd y^i)^2 + f \Big( N(x,y)^2 \, \dd x^2 + \gamma_{ij} \, \dd y^i \, \dd y^j \Big).
\end{equation}
These are smooth on $\cM^\pm$ but become singular on $\cS$. The signature of $h$ is $\sig{4}{0}$ if $f>0$ and $\sig{0}{4}$ if $f<0$. Hence in order to satisfy condition 2 of our definition \ref{ambipolar def} we need to show that $f$ changes sign at $\cS$. This will be true if $f$ has a {\it first order} zero on $\cS$. This can be motivated either by appealing to genericity, or (as we will show below and in \aref{app:general}) by using the equation of motion for the Maxwell field. If $f$ has a first order zero then we can choose the overall sign of our coordinate $x$ so that $f>0$ for $x>0$ and $f<0$ for $x<0$ so $h$ has signature $\sig{4}{0}$ in $\cM^+$ and signature $\sig{0}{4}$ in $\cM^-$.

We can now explain why we chose our coordinate $x$ to satisfy \eqref{xeq}. The reason is that these conditions imply that $x$ is harmonic w.r.t. the metrics $h$ on $\cM^\pm$, i.e.,
\be
\label{x harmonic4}
\dd \hodge_4 \dd x = 0
\ee 
This will be important when we use the AJS construction to construct the base space from the data on $\cS$.\footnote{
From the 4d perspective it is not obvious that there exist solutions of \eqref{x harmonic4} that vanish on $\cS$ because $h^\pm$ is singular on $\cS$. Our 5d definition of $x$ shows that such a solution does indeed exist for the class of base spaces arising from the 5d spacetimes under consideration here.}
  
We are free to choose an orthogonal basis for the 3-metric $\gamma_{ij}$. We will choose one of the basis 1-forms to be $\nu_i dy^i$ and write
\begin{equation}
\gamma_{ij} \, \dd y^i \, \dd y^j = (\rho^1)^2 + (\rho^2)^2 + Q(x,y) \, (\nu_i \, \dd y^i)^2,
\end{equation}
for some function $Q$.  Since $\gamma_{ij} \, \dd y^i \, \dd y^j$ is smooth and non-degenerate at $\cS$, this implies that we can choose $\rho^1, \rho^2$ that are smooth and non-vanishing on $\cS$ and hence they are smooth 1-forms on $\cM$. 
The base space metric on $\cM^\pm$ is then
\begin{equation} \label{sugra 4d metric}
h = f^{-1} (1+Q f^2) \, \nu^2 + f N^2 \, \dd x^2 + f \, \delta_{ab} \, \rho^a \, \rho^b.
\end{equation}
Next we consider the 2-forms $X^i$. On $\cM^\pm$ these are orthonormal and anti-self-dual with respect to the volume form
\begin{equation}
\label{volform}
\vol_h = (1+Qf^2)^{1/2} f N \, \rho^1 \wedge \rho^2 \wedge \nu \wedge \dd x.
\end{equation}
It is convenient to introduce an orthonormal basis of anti-self dual 2-forms:
\begin{align}
\Omega^a &= f N \, \dd x \wedge \rho^a + (1 + Qf^2)^{1/2} \, \varepsilon^{ab} \rho^b \wedge \nu, \\
\Omega^3 &= N (1+Qf^2)^{1/2} \, \dd x \wedge \nu + f \, \rho^1 \wedge \rho^2,
\end{align}
These satisfy the algebra \eqref{4dalg} on $\cM^\pm$. The 2-forms $X^i$ must be related to the 2-forms $\Omega^i$ by an $SO(3)$ rotation:
\be
 \label{Xijdef}
 X^i = X^i{}_j \Omega^j
\ee
for some $SO(3)$ matrix $X^i{}_j$. 

Note that the 2-forms $\Omega^i$ are smooth at $\cS$ and hence they can be regarded as smooth 2-forms on $\cM$. Since the $X^i$ are also smooth, it follows that the matrix $X^i{}_j$ must also be smooth at $\cS$.\footnote{In more detail: smoothness of the $\rho^a \wedge \nu$ components of $X^i$ implies that $X^i{}_a$ is smooth and smoothness of the $dx \wedge \nu$ component implies that $X^i{}_3$ is smooth.}

When we pull-back to $\cS$ we obtain
\begin{equation}
\cS^* \Omega^a =\varepsilon^{ab} \, \rho^b \wedge \nu|_{x=0} \neq 0, \qquad \cS^* \Omega^3 = 0,
\end{equation}
where the first pullback is nonzero because the basis $\rho^1, \rho^2, \nu$ is non-degenerate at $x=0$. We see that $\Span\{\cS^* \Omega^1, \cS^* \Omega^2, \cS^* \Omega^3\}$ is 2-dimensional. Since the $\Omega^i$ are related to $X^i$ by an $SO(3)$ rotation, it follows that $\cW \equiv \Span\{\cS^* X^1, \cS^* X^2, \cS^* X^3\}$ coincides with $\Span\{\cS^* \Omega^1, \cS^* \Omega^2, \cS^* \Omega^3\}$ and hence $\cW$ is two-dimensional, in agreement with condition 3(a) of our definition \ref{ambipolar def}. Since $\cS^* \Omega^a$ provide a basis for $\cW$ we can determine the distribution $\cD$ by taking the sum of their kernels. The result is that $\cD$ is the space of vectors on $\cS$ that is orthogonal to the 1-form $\cS^* \nu$. 
\subsubsection{Smoothness of the Maxwell 2-form $F$}

To satisfy condition 3(b) of our definition we must prove that $\cS^* \nu$ is a contact form on $\cS$. We will show that this is a consequence of smoothness of the Maxwell 2-form at $\cS$ in 5d, assuming that $f$ has a first order zero on $\cS$. We will then (in \aref{app:general}) use the Maxwell equation to justifiy this assumption. 

For $x \ne 0$ the Maxwell 2-form is given by \eqref{F omega} with $\omega=f^{-2} \nu$. Our strategy will be to write this in terms of the smooth basis $\{\dd t,\dd x,\nu,\rho^1,\rho^2\}$ and demand that the resulting expression can be smoothly extended across $x=0$. 

It will be useful to define a basis of vector fields $e_1, e_2, e_3$ dual to $\rho^1, \rho^2, \nu$:
\begin{equation}
\rho^a(e_b) = \delta^a{}_b, \qquad \nu(e_3) = 1, \qquad \rho^a(e_3) = 0, \qquad \nu(e_a) = 0.
\end{equation}
Then using
\begin{align}
\dd \omega &= -2 f^{-3} \, \dd f \wedge \nu + f^{-2} \, \dd \nu, \\
\dd \nu &\equiv (\dd \nu)_{x \nu} \, \dd x \wedge \nu + (\dd \nu)_{x a} \, \dd x \wedge \rho^a + (\dd \nu)_{a \nu} \, \rho^a \wedge \nu + (\dd \nu)_{12} \, \rho^1 \wedge \rho^2,
\end{align}
we obtain:
\begin{equation} \label{F expansion}
\begin{split}
2 \sqrt3 F &= -3 \, \dd t \wedge \dd f \\
& \qquad + \dd x \wedge \nu \, \Big[ - f^{-2} \partial_x f + 2 f^{-1} (\dd \nu)_{x \nu} + f^{-2} (1 + Q f^2)^{1/2} N \, (\dd \nu)_{12} \Big] \\
& \qquad + \rho^a \wedge \nu \, \Big[ - f^{-2} e_a(f) + 2 f^{-1} \, (\dd \nu)_{a \nu} - f^{-2} (1+Q f^2)^{1/2} N^{-1} \, \varepsilon^{ab} (\dd \nu)_{xb} \Big] \\
& \qquad + \dd x \wedge \rho^a \, \Big[ 2 f^{-1} \, (\dd \nu)_{xa} - 2 f^{-1} (1+Qf^2)^{-1/2} N \, \varepsilon^{ab} e_b(f) + (1+Qf^2)^{-1/2} N \, \varepsilon^{ab} (\dd \nu)_{b\nu} \Big] \\
& \qquad + \rho^1 \wedge \rho^2 \, \Big[ 2 f^{-1} \, (\dd \nu)_{12} - 2 f^{-1} (1+Qf^2)^{-1/2} N^{-1} \partial_x f + (1+Qf^2)^{-1/2} N^{-1} \, (\dd \nu)_{x \nu} \Big].
\end{split}
\end{equation}
Next we will expand this for small $x$ and demand that the singular terms vanish. To do this we must return to the question of how $f$ behaves at $x=0$. Smoothness of $f$ implies that we have $f = {\cal O}(x^p)$ for some positive integer $p$. 
In \aref{app:general}, we use the Maxwell equation \eqref{f eqn} to show that $p=1$, i.e., $f$ has a first order zero on $\cS$.\footnote{There is an apparent contradiction between this result and the results of \cite{Pasini:2015zlx}, wherein $f$ is contrived to vanish as $\cO(x^p)$ for $p$ arbitrarily high; however, in that paper, smoothness of the Maxwell field was not imposed.} 

We can now expand $f$ and $N$ as
\begin{equation}
\label{fNexp}
f(x, y) = x f_1(y) + x^2 f_2(y) + \cO(x^3), \qquad N(x,y) = N_0(y) + x N_1(y) + \cO(x^3).
\end{equation}
where $f_1$ and $N_0$ are non-zero. As discussed above, we can define $x$ so that $f>0$ for $x>0$ hence $f_1>0$. 

We see that there are singular parts of \eqref{F expansion} at orders $x^{-2}$ and $x^{-1}$. Requiring these to vanish implies  that $\dd \nu$ take the form\footnote{Similarly to before, the notation $\cO(x^2)(\dd x, \rho^a, x^{-1} \nu)$ denotes e.g. $\cO(x^2) \dd x \wedge \dd \rho^a$ or $\cO(x) \dd x \wedge \nu$ etc.}
\begin{equation} \label{nu contact}
\begin{split}
\dd \nu &= \big(\partial_x f -2 f (\dd \nu)_{x\nu} \big) N^{-1} \, \rho^1 \wedge \rho^2 + \big( e_a(f) -2 f (\dd \nu)_{a \nu} \big) N \, \varepsilon^{ab} \, \rho^b \wedge \dd x \\
& \qquad + (\dd \nu)_{x \nu} \, \dd x \wedge \nu + (\dd \nu)_{a \nu} \, \rho^a \wedge \nu + \cO(x^2)(\dd x, \rho^a, x^{-1} \nu),
\end{split}
\end{equation}
where the $\dd x \wedge \nu$ and $\rho^a \wedge \nu$ components are unconstrained. In particular, we see that
\begin{equation}
\cS^* \dd \nu = f_1 N_0^{-1} \, \cS^* (\rho^1 \wedge \rho^2) + \cS^* \big( (\dd \nu)_{a \nu} \, \rho^a \wedge \nu \big),
\end{equation}
hence $\cS^* (\nu \wedge \dd \nu) \neq 0$ and so $\cS^* \nu$ is a contact 1-form on $\cS$. Thus we have shown that $\cM$ together with the 2-forms $X^i$ satisfies all the conditions of our definition \ref{ambipolar def}. In summary, we have shown that a necessary condition for a 5d supersymmetric solution to have an evanescent ergosurface is that its base space be an ambipolar hyperk\"ahler space according to our definition.

\subsubsection{Comparison with \sref{construction}}

The coordinate $x$ introduced above is harmonic on the base space. Hence it must be possible to write our base space in the form determined in section \sref{construction}. We can compare directly the metrics \eqref{h defa} and \eqref{sugra 4d metric}. In both cases, $x$ is a harmonic coordinate on the base space. However, in \eqref{h defa}, $x$ is completely determined whereas in \eqref{sugra 4d metric} there is still some gauge freedom in $x$ arising from the freedom to choose $N_0$ (or $\alpha$). We can fix this freedom by comparing the $\dd x^2$ terms:
\begin{equation}
f N^2 = \varphi= x \big( 1 + 2 x \mu^a{}_a \big) + \cO(x^3),
\end{equation}
and hence the appropriate gauge choice for comparing with \sref{construction} is
\begin{equation}
N_0 = f_1^{-1/2}.
\end{equation}
We now compare other components of the metrics \eqref{h defa} and \eqref{sugra 4d metric}. By taking the norm of $\eta_3$ using both metrics we learn that 
\be
 h_{33} = f^{-1} \nu(\eta_3)^2 + {\cal O}(x) \qquad \Rightarrow \qquad \nu(\eta_3)^2 = \frac{f}{x} + {\cal O}(x^2)
\ee
where we used \eqref{hexp} in the second equality. By taking the inner product of $\eta_3$ and $\eta_a$ w.r.t.  both metrics we learn that
\be
 \cO(x^2) = f^{-1} \nu(\eta_3) \nu(\eta_a) + \cO(x) \qquad  \Rightarrow \qquad \nu(\eta_a) = \cO(x^2)
\ee
hence we must have
\be \label{nuexp}
 \nu = \left[ \left(\frac{f}{x}\right)^{1/2} +\cO(x^2)\right] \theta^3 + \cO(x^2)  \theta^a=f_1^{1/2} \left( 1 + \frac{xf_2}{2f_1} + \cO(x^2) \right) \theta^3 + \cO(x^2) \theta^a
\ee
where we used the freedom $\theta^3 \rightarrow -\theta^3$, $\theta^1 \leftrightarrow \theta^2$ to fix the sign in the first term. From this we obtain the behaviour of $\omega$ near $\cS$:
\be
\label{omegaexp}
 \omega = \left( \frac{1}{x^2 f_1^{3/2}} - \frac{3 f_2}{2x f_1^{5/2}} + \cO(1) \right) \theta^3 + \cO(1) \theta^a
\ee
In addition, by comparing \eqref{volform} and \eqref{vdef}, we can determine the volume form $v$,
\begin{equation} \label{ajs v}
v = (1+Qf^2)^{1/2} N^{-1} \, \rho^1 \wedge \rho^2 \wedge \nu,
\end{equation}
which is independent of $x$ as a consequence of \eqref{N0}. 

We can relate some of the other quantities used above to those of \sref{construction}. In \sref{construction} we denoted $\cS^* X^i$ as $Y^i$. From \eqref{Yibasis} and \eqref{Xijdef} we obtain
\be
\label{compare}
 t_i^a \epsilon^{ab} \theta^b \wedge \theta^3 = \cS^* \left( X^i{}_j \epsilon^{ab} \rho^b \wedge \nu  \right)
\ee
We identify
\be
\label{compare2}
 t_i^a = X^i{}_a |_{x=0} \qquad u_i = X^i{}_3|_{x=0}
\ee
which satisfy the algebraic relations \eqref{tnorm} and \eqref{usign} because $X^i{}_j$ is an $SO(3)$ matrix. From \eqref{compare} we must now have\footnote{Of course $\theta^a$ and $\rho^a$ are only defined up to $SO(2)$ rotations. We have made a particular choice in \eqref{compare2}.}
\be
 \theta^a = f_1^{1/2} \cS^* \rho^a + \beta^a \cS^* \nu
\ee
for some $\beta^a$. The $\beta^a$ are uniquely determined by the condition $ \dd \theta^3 = \theta^1 \wedge \theta^2$
which gives
\be
\theta^a = f_1^{1/2} \, \cS^* \rho^a - f_1^{-1} \, \varepsilon^{ab} (\dd \nu)_{b\nu} \, \cS^* \nu +\frac12 f_1^{-2} \, \varepsilon^{ab} \, \cS^* \big( e_b(f_1) \, \nu \big).
\ee

\subsubsection{Sufficient conditions for smoothness}

We have shown that necessary conditions for smoothness of a 5d supersymmetric configuration with an evanescent ergosurface is that the base space be an ambipolar hyperk\"ahler space, that $f$ behaves as in \eqref{fNexp} with $f_1 >0$ (appealing to genericity or the Maxwell equation), and that $\omega$ behave as in \eqref{omegaexp}. These conditions are also {\it sufficient} for smoothness. To see this, we plug the expansions \eqref{h def}, \eqref{fNexp} and \eqref{omegaexp} into \eqref{5d metric} to obtain the expansion of the 5d metric as\footnote{Again, the error term $\cO(x^2)(x \, \dd t, \dd x, \theta^a, x^{-1} \theta^3)$ means a quadratic form built out of the listed elements.}
\begin{equation}
\begin{split}
\label{5dmetricexp}
g &= - x^2 f_1^2 (1 + 2x f_1^{-1} f_2) \, \dd t^2 - 2 f_1^{1/2} \Big(1 + \frac12 x f_1^{-1} f_2 \Big) \, \dd t \, \theta^3 + f_1^{-1} (1 - x f_1^{-1} f_2) (1 + 2 x \mu^a{}_a) \, \dd x^2 \\
& \qquad + f_1^{-1} (1 - x f_1^{-1} f_2) \, \delta_{ab} (\theta^a + x \mu_{ac} \, \theta^c) (\theta^b + x \mu_{bd} \, \theta^d) + \cO(x^2)(x \, \dd t, \dd x, \theta^a, x^{-1} \theta^3),
\end{split}
\end{equation}
which is smooth at $x=0$ with Lorentzian signature. Furthermore, $x=0$ is a timelike hypersurface on which $f$ vanishes, i.e., an evanescent ergosurface. 

Note that the $(\theta^3)^2$ component of the metric is smooth at $x=0$ but we cannot determine its sign without taking the expansion of the base space metric to one order higher than we have done. It is possible that the sign of this component might be negative, in which case $\eta_3$ would be timelike w.r.t. the 5d metric. If $\eta_3$ has closed orbits then these would be closed timelike curves. So our construction does not guarantee freedom from closed timelike curves.

The expansions \eqref{h def}, \eqref{fNexp} and \eqref{omegaexp} determine the behaviour of $G^+$ as
\be
\label{Gplusexp}
 G^+ =  \frac{3}{2} \, \dd \left( \frac{\theta^3}{x f_1^{1/2}} \right) + \cO(1).
\ee
where $\cO(1)$ denotes terms smooth at $x=0$. The term in brackets here is the same as the singular part of $f\omega$. Hence the singular part of $(2/3) G^+$ cancels the singular part of $\dd(f \omega)$  in the expression \eqref{F def} for the Maxwell field. Therefore the Maxwell field is also smooth at $x=0$. 

In summmary, we have shown the following:

{\it The necessary and sufficient conditions for a configuration $(g,F)$ of 5d minimal supergravity to be smooth, with an evanescent ergosurface at which $f$ has a first order zero,\footnote{We repeat that the assumption of a first order zero can be justified by appealing to the equation of motion for $F$. But here we are stating our result in a way that refers only to supersymmetry and does not use the equations of motion explicity.} and admit a supercovariantly constant spinor, is that, when decomposed into the canonical form \eqref{5d metric}, the resulting base space is an ambipolar hyperk\"ahler manifold (with metric \eqref{h defa}) and the 1-form $\omega$ satisfies \eqref{omegaexp}. }

\section{Initial value construction of supersymmetric solutions}
\label{sec:ashtekar}

\subsection{Introduction}

The result just summarized concerns configurations of 5d minimal supergravity that admit a supercovariantly constant spinor. Now we want to ask whether any further restrictions emerge from demanding that the configuration is a solution of the equations of motion, i.e., that it satisfies the BPS equations.
In particular, given an arbitrary ambipolar hyperk\"ahler space $\cM$, can one use it to construct a 5d supersymmetric {\it solution} with an evanescent ergosurface without further restrictions on $\cM$? We will prove that the answer is yes, at least for the class of ambipolar hyperk\"ahler spaces that can be constructed using the method of \sref{construction} (which may well be all such spaces). 

The idea is to extend the `initial value' construction of the base space to an `initial value' construction of a solution of the BPS equations. In the AJS construction, initial data prescribed on a 3d manifold $\cS$ is used to construct a hyperk\"ahler manifold  containing $\cS$ as a hypersurface \cite{Ashtekar:1987qx}. We will show that prescribing additional data on $\cS$ allows us to solve the BPS equations on this manifold and thereby construct a 5d supersymmetric solution. We will do this both for the case for which $\cS$ is a regular hypersurface in a hyperk\"ahler manifold and the case for which $\cS$ is the privileged hypersurface of an ambipolar hyperk\"ahler manifold. In both cases, $\cS$ corresponds to a timelike hypersurface in the 5d spacetime. In the latter case, this hypersurface is an evanescent ergosurface. 

This is to be contrasted with the usual initial value problem in GR in which data is specified on a spacelike hypersurface and evolved in time. We are instead specifying data on a timelike hypersurface and evolving in a spacelike direction. This is usually an ill-posed problem. However, we are restricting ourselves to supersymmetric solutions, which are stationary and therefore expected to be analytic. Therefore one can hope that local existence and uniqueness of  solutions near $\cS$ can be established using the Cauchy-Kowalevski theorem. Of course it is difficult to discuss global regularity of solutions constructed this way but exactly the same remark applies to the AJS construction. 

In this section we will show that an AJS-like construction can indeed be developed for solving the BPS equations. This is straightforward when $\cS$ is a regular hypersurface in a hyperk\"ahler base space and, with some care, can also be done when $\cS$ is the privileged hypersurface of an ambipolar hyperk\"ahler base space, corresponding to an evanescent ergosurface. In both cases, we find that the initial data is equivalent to specifying 8 free functions on $\cS$, so the presence of an evanescent ergosurface does not impose functional constraints on the solution. 

\subsection{Initial data on a regular hypersurface}

In this section we will show how to solve the BPS equations starting from initial data prescribed on a smooth, oriented, 3d manifold $\cS$ for the case in which $\cS$ is a regular hypersurface within a hyperk\"ahler space $\cM$. 

The essence of the AJS `initial value' construction of hyperk\"ahler manifolds is that it distills the hyperk\"ahler problem into an evolution problem for a collection of vector fields $V_i$.  Therefore we seek to mimic this method for the BPS equations, by expressing quantities in terms of such vector fields as much as possible. 

We will solve the BPS equations in the order suggested by  \cite{Bena:2004de}. The novelty here is that we will formulate each equation as an initial value problem.

We follow the notation of \aref{app:ashtekar}: we assume that we have constructed a hyperk\"ahler space on the manifold $\cM = (-\epsilon,\epsilon) \times \cS$ with coordinates $(x,y^i)$ where $x \in (-\epsilon,\epsilon)$ and $y^i$ are coordinates on $\cS$. The hypersurface $\cS$ is identified with the surface $x=0$ in $\cM$. 

We start with the equation \eqref{Gplus}. We can express the self-dual 2-form $G^+$ in terms of a vector field $W$ tangent to the level sets of $x$ via the formula 
\begin{equation}
\label{Wdef}
G^+ = \dd x \wedge \ushort{W} - \ins_W v,
\end{equation}
where $v$ is the volume 3-form of the AJS construction, and $\ushort W \equiv h(W)$ is the 1-form obtained by `lowering an index' on $W$ with the hyperk\"ahler metric. We now impose $\dd G^+ = 0$, which gives
\begin{equation}
\label{Wconstraint}
\ddh (\ins_W v) = 0
\end{equation}
\begin{equation}
\label{Wevolution}
\ddh \ushort W + \ins_{\partial_x W} v = 0.
\end{equation}
where $\ddh$ denotes the pull-back of the exterior derivative to the level-sets of $x$ (i.e. in coordinates $(x,y^i)$ it involves differentiation only w.r.t. $y^i$). Equation \eqref{Wevolution} uniquely determines $\partial_x W$, i.e., it is a first-order evolution equation for $W$. The identity $\dd \dd G^+=0$ implies that, when this evolution equation is satisfied, we automatically have
\be
\label{constraintpreserved}
 \partial_x \ddh (\ins_W v)= 0
\ee
and hence if the divergence-free constraint \eqref{Wconstraint} is satisfied on $\cS$ then it is satisfied everywhere. Therefore we can solve the equation $\dd G^+ =0$ by specifying a divergence-free vector field $W$ on $\cS$ and then using this as the initial condition to solve \eqref{Wevolution}. Such initial data contains $2$ free functions, i.e., $1$ 4d degree of freedom, as expected if we regard $G^+$ as a self-dual solution of Maxwell's equations in 4d. 

Now consider \eqref{f eqn} which is a standard Poisson equation for $f^{-1}$. In terms of the AJS vector fields $V_i$ we find
\begin{equation}
\dd \hodge_4 \dd f^{-1} = \Big( \partial_x^2 (f^{-1}) + V_i (V_i (f^{-1})) \Big) v \wedge \dd x.
\end{equation}
We also find
\begin{equation}
G^+ \wedge G^+ = 2 \, \ushort W \wedge (\ins_W v) \wedge \dd x = 2 \, h(W, W) \, v \wedge \dd x,
\end{equation}
and hence \eqref{f eqn} reduces to
\begin{equation} \label{f eqn explicit}
\partial_x^2 (f^{-1}) + V_i (V_i (f^{-1})) = \frac89 \, h(W,W).
\end{equation}
Local existence and uniqueness of solutions is guaranteed by the Cauchy-Kowalevski theorem if we prescribe $f$ and its normal derivative on $\cS$. The only restriction is $f|_\cS \ne 0$, which is to be expected since we are not considering an ambipolar base space here. In summary, we have to specify $2$ free functions on $\cS$ to solve \eqref{f eqn}, equivalent to $1$ degree of freedom in 4d. 

Finally, to solve \eqref{omega eqn} we express the anti-self dual 2-form $G^-$ in terms of another vector field $Z$ tangent to the level sets of $x$ via the formula 
\begin{equation}
\label{Zdef}
G^- = \dd x \wedge \ushort{Z} + \ins_Z v,
\end{equation}
We now take the exterior derivative of \eqref{omega eqn} to obtain 
\be
\label{ddomega}
 \dd \left[ f^{-1}(G^+ + G^-) \right]=0
\ee
which gives an evolution equation for $Z$ 
\begin{equation} \label{Z evolution}
0 =  \ins_{\partial_x Z} v + f^{-1} \, \ddh f \wedge (\ushort W + \ushort Z) + f^{-1} \partial_x f \, (\ins_W v - \ins_Z v) - \ddh \ushort Z,
\end{equation}
together with a constraint
\begin{equation} \label{Z constraint}
0 = \ddh \big( \ins_Z v \big) + f^{-1} \Big( W(f) - Z(f) \Big) \, v.
\end{equation}
Similar to the case of $G^+$, one can show that the constraint \eqref{Z constraint} is automatically preserved by the evolution equation \eqref{Z evolution}.  Thus the initial condition for $Z$ consists of one vector field on $\cS$ satisfying one constraint on its divergence, leaving a total of 2 free functions on $\cS$, i.e., one 4d degree of freedom. 

Having solved for $G^-$ we now substitute it into the RHS of \eqref{omega eqn}. This determines $\omega$ up to an exact differential $\dd a$ for some function $a$. The latter is a gauge degree of freedom that can be eliminated by a shift in the time coordinate: $t \rightarrow t+a$. 

We have shown how a supersymmetric 5d solution can be constructed from initial data on $\cS$ for the case of a regular hyperk\"ahler base space. We can count the degrees of freedom from this prescription.  First, the hyperk\"ahler base is determined by 2 gauge invariant functions on $\cS$ (see \aref{app:ashtekar}).  The 2-forms $G^\pm$ and the function $f$ are each determined by 2 more free functions on $\cS$. This gives a total of 8 functions on $\cS$. This is equivalent to $4$ degrees of freedom in 4d. 

\subsection{Ambipolar base space}

We now want to investigate whether we can formulate the BPS equations as an initial value problem starting from the canonical surface $\cS$ in an ambipolar hyperk\"ahler base space, and solve so that the resulting 5d solution is smooth with an evanescent ergosurface. We will show that this can indeed be done for the class of ambipolar hyperk\"ahler spaces that can be constructed (locally) using the method of \sref{construction}. The difference from the previous section is that there we assumed all quantities were smooth at $\cS$. However, we must now deal with the fact that the base space metric $h$ and the 1-form $\omega$ (and hence also $G^\pm$) are singular at $\cS$, i.e., we are trying to solve an initial value problem where some quantities are singular at the initial surface. Nevertheless, since we have already determined the nature of this singular behaviour, it will turn out that we can indeed solve each equation as an initial value problem. 

We assume that our base space is an ambipolar hyperk\"ahler manifold constructed using the method of \sref{construction}. The base space metric is given by \eqref{h defa}. Expanding in components we have
\begin{equation} \label{h generic}
h = h_{33} \, (\theta^3)^2 + h_{3a} \big( \theta^3 \otimes \theta^a + \theta^a \otimes \theta^3 \big) + h_{ab} \, \theta^a \otimes \theta^b + h_{xx} \, \dd x^2.
\end{equation}
We will expand each component as a series in $x$,
\begin{equation} \label{h expansion}
h_{33} = x^{-1} h_{33}^{(-1)} + \sum_{k=0}^\infty x^k h_{33}^{(k)}, \qquad h_{3a} = \sum_{k=1}^\infty x^k h_{3a}^{(k)}, \qquad h_{ab} = \sum_{k=1}^\infty x^k h_{ab}^{(k)},
\end{equation}
where the coefficients are functions of the $y^i$.  Comparing to \eqref{h def}, we see in particular that 
\be
\label{h33}
h_{33}^{(-1)} = 1 \qquad h_{33}^{(0)} = 0
\ee
We will also require that $f$ is expanded as in \eqref{fNexp} and that $\omega$ has the behaviour \eqref{omegaexp}, since we know these are necessary (and sufficient) for smoothness of the 5d fields.

\subsubsection{The $G^+$ equation}

The behaviour of $G^+$ near $\cS$ required for 5d smoothness was determined in \eqref{Gplusexp} where $f_1$ is a positive function. Expanding this gives
\begin{equation}
\label{Gplusexpcomp}
G^+ = -\frac{3}{2f_1^{1/2} x^2} \, \dd x \wedge \theta^3 -\frac{3}{4 x f_1^{3/2}} \, \ddh f_1 \wedge \theta^3 + \frac{3}{2x f_1^{1/2}} \, \theta^1 \wedge \theta^2 + \cO(1)
\end{equation}
To solve the $G^+$ equation \eqref{Gplus}, we first write $G^+$ in terms of a vector field $W$ tangent to the level sets of $x$, as in \eqref{Wdef} so that the $G^+$ equation becomes \eqref{Wconstraint} and \eqref{Wevolution}. From the above behaviour of $G^+$ we see that $W$ is $\cO(x^{-1})$ as $x \rightarrow 0$. In components we have
\begin{gather}
 W = W^a \, \eta_a + W^3 \, \eta_3, \\
\ushort W = \big( h_{ab} W^b + h_{3a} W^3 \big) \, \theta^a + \big( h_{3a} W^a + h_{33} W^3 \big) \, \theta^3.
\end{gather}
The evolution equation \eqref{Wevolution} can be expanded in components,
\begin{equation}
\begin{split}
0 &= \theta^1 \wedge \theta^2 \, \Big[ \partial_x W^3 + \varepsilon^{ca} \eta_c \big( h_{ab} W^b + h_{3a} W^3 \big) - \varepsilon^{ac} \kappa_c \big( h_{ab} W^b + h_{3a} W^3 \big) \\
& \qquad \qquad \qquad \qquad + f_1 N_0^2 \big( h_{3a} W^a + h_{33} W^3 \big) \Big] \\
& \qquad + \theta^b \wedge \theta^3 \, \Big[ \varepsilon^{ab} (\partial_x W^a) - \eta_3 \big( h_{ab} W^a + h_{3b} W^3 \big) + \eta_b \big(h_{3a} W^a + h_{33} W^3 \big) \\
& \qquad \qquad \qquad \qquad + \lambda^a{}_b \big( h_{ac} W^c + h_{3a} W^3 \big) \Big],
\end{split} \label{Gplus series}
\end{equation}
where $\kappa_a, \lambda^a{}_b$ are defined in \eqref{eta alg} and \eqref{kappas}.  

If the evolution equation is satisfied then we have \eqref{constraintpreserved} as above and hence
\be
 \ddh (\ins_W v) = \chi
\ee
where $\partial_x \chi=0$. Equation \eqref{Wconstraint} is now equivalent to the condition $\chi=0$ on $\cS$. However, since $W$ is singular on $\cS$ it is not immediately obvious how to arrange $\chi=0$ on $\cS$. To investigate this we set
\be
\label{Wtildedef}
 W = \frac{1}{x} \widetilde{W}
\ee
where $\widetilde{W}$ is smooth on $\cS$. We then have
\be
\label{Wtilde}
 \ddh (\ins_{\widetilde W} v) = x \chi
\ee
and hence
 \be
 \ddh (\ins_{\partial_x \widetilde W} v) = \chi
\ee
The LHS is now smooth on $\cS$. Hence the constraint $\chi=0$ reduces to
\be
 \ddh (\ins_{\partial_x \widetilde W} v)|_{\cS} = 0,
\ee
In other words, when the evolution equation is satisfied, \eqref{Wconstraint} will also be satisfied iff $\partial_x \widetilde{W}$ is divergence-free on $\cS$.

Next we write series expansions in $x$ for the components of $W$,
\begin{equation}
\label{Wseries}
W^3 = x^{-1} W^3_{(-1)} + \sum_{k=0}^\infty x^k W^3_{(k)}, \qquad W^a = x^{-1} W^a_{(-1)} + \sum_{k=0}^\infty x^k W^a_{(k)}.
\end{equation}
where $W^3_{(k)}$  and $W^a_{(k)}$ are functions on $\cS$. Matching to \eqref{Gplusexpcomp} determines the singular terms:
\be \label{W-1}
 W^3_{(-1)} = - \frac32 f_1^{-1/2}, \qquad W^a_{(-1)} = - \frac32 \varepsilon^{ab} \eta_b \big( f_1^{-1/2} \big),
\ee
We have
\be
 \partial_x \widetilde{W}|_{\cS} = W^a_{(0)} \eta_a + W^3_{(0)} \eta_3
\ee
so our constraint equation will be satisfied iff the RHS here is divergence-free.

Plugging the expansion into \eqref{Gplus series}, we can extract the lowest few powers of $x$.  First,  the $\theta^1 \wedge \theta^2$ component:
\begin{align}
x^{-2} &: \quad 0 = - W^3_{(-1)} + h_{33}^{(-1)} W^3_{(-1)}, \label{W3-1} \\
x^{-1} &: \quad 0 = h_{33}^{(-1)} W^3_{(0)} + h_{33}^{(0)} W^3_{(-1)}, \\
\begin{split}
x^0 &: \quad 0 = W^3_{(1)} + h_{33}^{(-1)} W^3_{(1)} + h_{33}^{(1)} W^3_{(-1)} + h_{33}^{(0)} W^3_{(0)} \\
& \qquad \qquad + \varepsilon^{ca} \eta_c \big(h_{ab}^{(1)} W^b_{(-1)} + h_{3a}^{(1)} W^3_{(-1)} \big) + h_{3a}^{(1)} W^a_{(-1)} \\
& \qquad \qquad - \varepsilon^{ac} \kappa_c \big(h_{ab}^{(1)} W^b_{(-1)} + h_{3a}^{(1)} W^3_{(-1)} \big),
\end{split} \label{W31}
\end{align}
using \eqref{h33} we see that the $\cO(x^{-2})$ terms cancel automatically and the $\cO(x^{-1})$ terms give
\begin{equation}
W^3_{(0)} = 0.
\end{equation}
The $\cO(x^0)$ terms then fix $W^3_{(1)}$ uniquely. Extending to higher orders we find that the $\cO(x^n)$ terms fix $W^3_{(n+1)}$ in terms of $W^3_{(k)}$ and $W^a_{(k)}$ with $k \le n$.

Now, expanding the $\theta^b \wedge \theta^3$ component, we get
\begin{align}
x^{-2} &: \quad 0 = - \varepsilon^{ab} W^a_{(-1)} + \eta_b \big(h_{33}^{(-1)} W^3_{(-1)} \big), \\
x^{-1} &: \quad 0 = \eta_b \big(h_{33}^{(-1)} W^3_{(0)} \big) + \eta_b \big( h_{33}^{(0)} W^3_{(-1)} \big), \\
\begin{split}
x^0 &: \quad 0 = \varepsilon^{ab} W^a_{(1)} + \eta_b \big( h_{33}^{(1)} W^3_{(-1)} + h_{33}^{(-1)} W^3_{(1)} \big) + \eta_b \big( h_{33}^{(0)} W^3_{(0)} \big) \\
& \qquad \qquad - \eta_3 \big(h_{ab}^{(1)} W^a_{(-1)} + h_{3b}^{(1)} W^3_{(-1)} \big) + \eta_b \big( h_{3a}^{(1)} W^a_{(-1)} \big) \\
& \qquad \qquad + \lambda^a{}_b \big(h_{ac}^{(1)} W^c_{(-1)} + h_{3a}^{(1)} W^3_{(-1)} \big),
\end{split}
\end{align}
Using \eqref{h33} and \eqref{W-1} we find that the $\cO(x^{-2})$ and $\cO(x^{-1})$ terms cancel automatically. The $\cO(x^0)$ terms fix $W^a_{(1)}$ uniquely. Extending to higher orders we find that the $\cO(x^n)$ terms fix $W^3_{(n+1)}$ in terms of $W^3_{(k+1)}$ and $W^a_{(k)}$ with $k \le n$.

It follows that we can solve the evolution equation recursively, order by order to determine all coefficients $W^3_{(n)}$ and $W^a_{(n)}$ {\it except} for $W^a_{(0)}$, which is therefore the initial data for the evolution equation. These two free functions are constrained by the condition that $W^a_{(0)} \eta_a$ must be divergence-free, leaving $1$ free function. However, we must not overlook the singular part of $W$, which is determined by $f_1$. It will be convenient to regard this (positive) free function as part of the initial data for $G^+$ rather than as initial data for $f$. Therefore we have shown that one can solve \ref{Gplus} to determine $G^+$ uniquely given initial data consisting of $2$ free functions on $\cS$.

\subsubsection{The $f$ equation}

Next we tackle the $f$ equation \eqref{f eqn explicit}:
\begin{equation} \label{f eqn explicit 2}
\partial_x^2 (f^{-1}) + V_i (V_i (f^{-1})) = \frac89 \, h(W,W).
\end{equation}
To find a solution, it is more convenient to expand $f^{-1}$, rather than $f$ itself, as a series in $x$.  Put
\begin{equation}
f^{-1} = x^{-1} q_{(-1)} + \sum_{k=0}^\infty x^k q_{(k)}, \qquad V_i = \sum_{k=0}^\infty x^k \, V_i^{(k)}.
\end{equation}
where the $q_{(k)}$ are functions on $\cS$ with
\be
q_{(-1)} = f_1^{-1}.
\ee
The left-hand side of the $f$ equation becomes
\begin{equation} \label{f eqn LHS}
\begin{split}
\partial_x^2 (f^{-1}) &+ (V_i (V_i (f^{-1})) \\
&= 2 x^{-3} q_{(-1)} + x^{-1} \big( V_i^{(0)} \big( V_i^{(0)} \big( q_{(-1)} \big) \big) \\
& \qquad + \sum_{k=0}^\infty x^k \Big[ (k+2)(k+1) q_{(k+2)} + \sum_{\ell = 0}^{k+1} \big( V_i^{(k - \ell + 1)} \big( V_i^{(\ell)}  \big( q_{(-1)} \big) \big) \\
& \qquad \qquad \qquad + \sum_{m=0}^k \sum_{\ell=0}^{k-m} \big( V_i^{(m)} \big( V_i^{(k-m-\ell)} \big( q_{(\ell)} \big) \big) \Big].
\end{split}
\end{equation}
The source term on the RHS is
\begin{equation} \label{f eqn RHS}
\begin{split}
\frac89 h(W,W) &= \frac89 x^{-3} \Big[ h_{33}^{(-1)} (W^3_{(-1)})^2 \Big] \\
& \quad + \frac89 x^{-1} \Big[ 2 h_{33}^{(-1)} W^3_{(-1)} W^3_{(1)} + h_{33}^{(1)} (W^3_{(-1)})^2 + 2 h_{3a}^{(1)} W^3_{(-1)} W^a_{(-1)} + h_{ab}^{(1)} W^a_{(-1)} W^b_{(-1)} \Big] \\
& \quad + \cO(1),
\end{split}
\end{equation}
A calculation shows that the singular terms cancel between \eqref{f eqn LHS} and \eqref{f eqn RHS} upon plugging in \eqref{h33}, \eqref{W-1}, and applying \eqref{W31}. At $\cO(x^n)$, $n \ge 0$, we obtain a recursion relation relating $q_{(n+2)}$ to $q_{(k)}$ with $k \le n$. Hence the solution is uniquely determined once we have specified $q_{(0)}$ and $q_{(1)}$, which are free data. So we have shown that \eqref{f eqn} can be solved to determine $f$ uniquely given initial data on $\cS$ consisting of $2$ more free functions.

\subsubsection{The $\omega$ equation}

Finally, we address the $\omega$ equation \eqref{omega eqn}. As before, we first write $G^-$ in terms of a vector field $Z$ tangent to the level sets of $x$, as in \eqref{Zdef} and use \eqref{ddomega} to obtain the evolution equation \eqref{Z evolution} and the constraint \eqref{Z constraint}. 

The evolution equation has a $\theta^1 \wedge \theta^2$ component:
\begin{equation} \label{Z series 1}
\begin{split}
0 &= \partial_x Z^3 + f^{-1} \partial_x f (W^3 - Z^3) - h_{3a} Z^a - h_{33} Z^3 \\
& \qquad + f^{-1} \, \varepsilon^{ca} \eta_c(f) \Big( h_{ab} W^b + h_{3a} W^3 + h_{ab} Z^b + h_{3a} Z^3 \Big) \\
& \qquad - \varepsilon^{ca} \eta_c \big( h_{ab} Z^b + h_{3a} Z^3 \big) + \varepsilon^{ac} \kappa_c \big( h_{ab} Z^b + h_{3a} Z^3 \big),
\end{split}
\end{equation}
and a $\theta^b \wedge \theta^3$ component:
\begin{equation} \label{Z series 2}
\begin{split}
0 &= \varepsilon^{ab} (\partial_x Z^a) + f^{-1} \partial_x f \, \varepsilon^{ab} (W^a - Z^a) - \lambda^a{}_b \big( h_{ac} Z^c + h_{3a} Z^3 \big) \\
& \qquad + f^{-1} \eta_b(f) \Big( h_{3a} W^a + h_{33} W^3 + h_{3a} Z^a + h_{33} Z^3 \Big) \\
& \qquad - f^{-1} \eta_3(f) \Big( h_{ab} W^a + h_{3b} W^3 + h_{ab} Z^a + h_{3b} Z^3 \Big) \\
& \qquad + \eta_3 \big( h_{ab} Z^a + h_{3b} Z^3 \big) - \eta_b \big( h_{3a} Z^a + h_{33} Z^3 \big).
\end{split}
\end{equation}
The behaviour of $Z$ near $\cS$ required for 5d smoothness can be determined using \eqref{omegaexp} together with the expansions of $f$ and the base space metric. We find that this gives
\be
\label{Zsing}
 Z = \left( -\frac{1}{2x f_1^{1/2} } - \frac{f_2}{2 f_1^{3/2}} + \cO(x) \right) \eta_3 + \left( \frac{3}{2x} \epsilon^{ab} \eta_b(f_1^{-1/2}) + \cO(1) \right) \eta_a
\ee
and hence we can write
\be
 Z = \frac{1}{x} \widetilde{Z}
\ee
where $\widetilde{Z}$ is smooth at $\cS$. 

The constraint equation \eqref{Z constraint} involves quantities singular on $\cS$. This problem can be addressed in a way similar to what we did with the $G^+$ equation. Let
\begin{equation}
\psi = f^{-1} \ddh \big( \ins_Z v \big) + f^{-2} \Big( W(f) - Z(f) \Big) \, v.
\end{equation}
Taking an exterior derivative of \eqref{ddomega} shows that 
\be
 \partial_x \psi=0
\ee
provided that the evolution equation \eqref{Z evolution} is satisfied. In terms of quantities smooth at $x=0$ we have
\be
\label{xpsi}
 x^2 \tilde{f} \psi = \ddh (\ins_{ \widetilde Z} v) + \tilde f^{-1} \big( \widetilde W(\tilde f) - \widetilde Z(\tilde f) \big) \, v 
\ee
where the smooth quantity $\widetilde{W}$ was defined in \eqref{Wtildedef} and the smooth non-zero quantity $\tilde{f}$ is defined by
\be
 f = x \tilde{f}.
\ee
Taking {\it two} $x$-derivatives of \eqref{xpsi}, we see that $\psi$ vanishes at $x=0$, and hence vanishes everywhere, iff
\be
\label{Zconstr2}
\Big\{ \ddh (\ins_{\partial_x^2 \widetilde Z} v) + \partial_x^2 \Big[ \tilde f^{-1} \big( \widetilde W(\tilde f) - \widetilde Z(\tilde f) \big) \Big] \, v \Big\} \Big|_\cS =0
\ee
Hence if the initial data for $Z$ satisfies this constraint then the evolution equation \eqref{Z evolution} guarantees that the constraint \eqref{Z constraint} is satisfied everywhere. 

We now expand as a series in $x$
\begin{equation}
Z^3 = x^{-1} Z^3_{(-1)} + \sum_{k=0}^\infty x^k Z^3_{(k)}, \qquad Z^a = x^{-1} Z^a_{(-1)} + \sum_{k=0}^\infty x^k Z^a_{(k)}.
\end{equation}
where, from \eqref{Zsing} we have
\be
 Z^3_{(-1)} = - \frac12 f_1^{-1/2}, \qquad Z^3_{(0)} = - \frac12 f_1^{-3/2} f_2, \qquad Z^a_{(-1)} = \frac32 \varepsilon^{ab} \eta_b \big( f_1^{-1/2} \big)
\ee
In this case, one must tediously carry out the expansion of equations \eqref{Z series 1}, \eqref{Z series 2}  to \emph{three} orders in $x$ in order to find the free functions in the $Z$ expansions. The result of this analysis is that
\be
Z^a_{(0)} = W^a_{(0)} - \frac32 f_1^{-3/2} \, \varepsilon^{ab} \eta_b \big( f_2 \big) - \frac92 f_1^{-1} f_2 \, \varepsilon^{ab} \eta_b \big( f_1^{-1/2} \big),
\ee
that  $Z^3_{(1)}$ is determined uniquely in terms of the $Z^3_{(n)}, Z^a_{(n)}$ for $n<1$; but that $Z^a_{(1)}$ and $Z^3_{(2)}$ are unconstrained by the evolution equation. Once these functions are specified, the evolution equation determines all higher order terms in the expansion. For example, examining the $\cO(x^1)$ part of \eqref{Z series 2} determines uniquely $Z^a_{(2)}$.  Hence  $Z^a_{(1)}$ and $Z^3_{(2)}$ are the initial data required to determine $G^-$.

We now consider \eqref{Zconstr2}, which depends on $Z^3_{(n)}$ and $Z^a_{(n)}$ for $n \le 1$. Hence this equation imposes one constraint on $Z^a_{(1)}$ in terms of known quantities. Hence specifying $Z^a_{(1)}$ is equivalent to specifying $1$ free function on $\cS$. $Z^3_{(2)}$ is unconstrained so $G^-$ is uniquely determined from initial data consisting of two free functions on $\cS$.

Now we know $G^\pm$, equation \eqref{omega eqn} determines $\omega$ up to an exact form $\dd a$, and we can eliminate $a$ via a shift of the time coordinates $t \rightarrow t+a$. Of course the resulting $\omega$ satisfies the condition \eqref{omegaexp} required for smoothness of the 5d solution because we used this condition to fix the behaviour of $G^\pm$ as $x \rightarrow 0$.

\subsubsection{Summary}

We have shown that we can solve the BPS equations as an initial value problem formulated in terms of data specified on the canonical surface $\cS$ of an ambipolar hyperk\"ahler manifold $\cM$. We saw previously that $\cM$ is determined by two free functions on $\cS$. We have just shown that two more free functions are required to determine each of $G^+$, $f$ and $G^-$ so the total number of free functions required to construct the 5d solution is $8$, equivalent to $4$ degrees of freedom in 4d. This is exactly as we found for the case in which $\cS$ was a regular surface within a hyperk\"ahler manifold. Note that no restrictions on $\cM$ emerged from this analysis so we have confirmed that it is possible to construct a smooth supersymmetric 5d solution with an evanescent ergosurface starting from any ambipolar hyperk\"ahler base space of the form constructed using the method of \sref{construction}. 

%
%

\section{Conclusions / Discussion}
\label{sec:conclusions}

In this paper we have defined the notion of an ambipolar hyperk\"ahler manifold, generalizing the notion of a folded hyperk\"ahler manifold of \cite{Hitchin:2015H,Biquard:2015cia}. Such manifolds first arose in the context of the `fuzzball' or microstate geometries program \cite{Giusto:2004kj,Bena:2005va,Berglund:2005vb,Bena:2007kg,Gibbons:2013tqa} as a curious way of side-stepping an inconvenient uniqueness theorem for hyperk\"ahler metrics.  By allowing metrics which change signature from $\sig{4}{0}$ to $\sig{0}{4}$, the assumptions of this theorem are violated. This feature is vitally important to the construction of large families of microstate geometries in 5-dimensional supergravity.

Evanescent ergosurfaces are a phenomenon which has been observed to be associated with the critical surfaces of ambipolar metrics \cite{Gibbons:2013tqa}.  On an evanescent ergosurface, a Killing vector which is asymptotically timelike becomes null, and then again timelike on the other side.  Thus unlike an ordinary ergosurface, an evanescent ergosurface is not the boundary of an ergoregion; it is more like an ergoregion that has been squished into a surface of zero thickness.

While it has been known that ambipolar hyperk\"ahler manifolds and evanescent ergosurfaces are associated with one another in 5d supergravity, neither a precise definition of such manifolds nor a precise explanation of this association, in full generality, has been written down in the literature.  In this paper we have supplied such a definition.  This definition encompasses all of the explicit ambipolar hyperk\"ahler metrics which can be written down within the Gibbons-Hawking ansatz, but also includes much more general metrics which cannot be written in this form.  Using methods analogous to \cite{Biquard:2015cia}, we employed the AJS construction \cite{Ashtekar:1987qx} to show that an ambipolar hyperk\"ahler manifold can be constructed from data specified on the critical surface $\cS$. 

We then considered the relation of such manifolds to solutions of 5d minimal supergravity with evanescent ergosurfaces.  We proved that a supersymmetric field configuration is smooth with an evanescent ergosurface if, and only if, the base space is an ambipolar hyperk\"ahler manifold (and the $1$-form $\omega$ behaves appropriately near $\cS$).  This result is interesting because it means that the signature flip from $\sig{4}{0}$ to $\sig{0}{4}$ cannot happen in an arbitrary way but only in the precise way specified by our definition. 

Finally, we showed how the entire 5-dimensional supergravity solution can be constructed uniquely from `initial' data specified on $\cS$. We did this first for the case in which $\cS$ is a surface within a regular hyperk\"ahler manifold and second for $\cS$ the canonical surface of an ambipolar hyperk\"ahler manifold. We found the same number of degrees of freedom are present in both cases. Therefore the presence of evanescent ergosurfaces does not place any functional constraints on the solution.  

We will now make some suggestions for future research. We proved that the 5d metric constructed from an ambipolar hyperk\"ahler base space (and appropriate $\omega$) is smooth with an evanesecent ergosurface. However, this metric may still be causally pathological in the sense that it may contain closed causal curves. Of course our construction is entirely local so we can't prove the absence of closed causal curves in general. But the usual way in which such curves show up in supersymmetric solutions would correspond to the $(\theta^3)^2$ component of the 5d metric \eqref{5dmetricexp} being non-positive at $x=0$, which would give closed causal curves if $\eta_3$ has closed orbits. It might be interesting to extend our expansions to higher order to determine the restriction on our initial data that results from demanding that this component be positive on $\cS$. (Since this restriction is an inequality it will not reduce the number of free functions on $\cS$.)

A straightforward generalization of our work would be to use the results of \cite{Gutowski:2004yv} to show that everything we have done can be extended to 5d minimal supergravity coupled to vector multiplets.

Our work can be placed into the context of a more general problem, namely understanding the nature of zero-sets of $f$. These can be classified by codimension. If the zero-set of $f$ has codimension $0$ then $f$ vanishes throughout some region. This is the `null class' of supersymmetric solutions classified in \cite{Gauntlett:2002nw}. Codimension $1$ corresponds to the case of $f$ vanishing on a hypersurface. This can be either null or timelike (because $K$ must be tangent to it). The null case is the case of a supersymmetric Killing horizon, which was analyzed in \cite{Reall:2002bh}. The timelike case is the case of an evanescent ergosurface considered in this paper. It would be interesting to investigate whether cases with higher codimension are possible. 

We have investigated only supersymmetric solutions to 5d supergravity.  Ambipolar-type effects also show up in non-supersymmetric solutions \cite{Goldstein:2008fq,Bena:2009fi,Bena:2009en,Bena:2013gma}, including some cases where the base space is K\"ahler but not hyperk\"ahler \cite{Bobev:2011kk,Niehoff:2013mla}.  It would be worth investigating how smoothness of the 5d structures imposes constraints on the critical surfaces of the 4d base space in these cases.  

Finally, we have discussed evanescent ergosurfaces only in 5 dimensions.  It would be interesting to investigate them for supersymmetric solutions of supergravity theories in other numbers of dimensions.

\section*{Acknowledgements}

We would like to thank N.~Hitchin, O.~Biquard, M.~Dunajski, and G.~W.~Gibbons for useful discussions, and N.~Bobev for drawing our attention to Ref. \cite{Bena:2007ju}.  This work was supported by ERC grant ERC-2011-StG 279363-HiDGR.


\section*{Appendices}

\appendix

\section{The AJS formalism for hyperk\"ahler metrics}
\label{app:ashtekar}

The Ashtekar-Jacobson-Smolin (AJS) formalism \cite{Ashtekar:1987qx} is a convenient way to formulate the problem of finding half-flat (or hyperk\"ahler) metrics as an initial value problem.  Here we give a brief exposition in notation suited to our application.

The premise is as follows:  Suppose we have a hyperk\"ahler metric $h$ with K\"ahler 2-forms $X^i$.  Choose a harmonic coordinate $x$ that vanishes on some hypersurface $\cS$. Let $y^i$ be coordinates on $\cS$ and extend them to a neighbourhood of $\cS$ by `carrying' them along the integral curves of $h^{-1}(dx)$. This gives a coordinate chart $(x,y^i)$ in a neighbourhood of $\cS$. Define a volume 3-form $v$ on $\cS$ and three vector fields $V_i$ via
\begin{equation}
\label{vdef}
v \wedge \dd x = \norm{\dd x}_h^2 \, \vol_h, \qquad V_i = (h^{-1} \circ X^i) \Big(\frac{\partial}{\partial x} \Big),
\end{equation}
here treating $h$ and $X^i$ as linear maps from $T \cM \to T^* \cM$.  One can then show that the following are true:
\begin{equation} 
\label{div-free}
\Lie_{V_i} v = 0
\end{equation}
\be
\label{nahm}
\partial_x V_i + \tfrac12 \varepsilon_{ijk} \, [V_j, V_k] = 0.
\ee
Thus, the $V_i$ are a set of divergence-free vector fields (with respect to $v$) on the level sets of $x$, which solve a set of first-order evolution equations (the Nahm equations) in $x$.  Since $x$ is harmonic, one finds
\begin{equation}
\Lie_{\partial_x} (v \wedge \dd x) = 0, \qquad \text{and hence} \qquad \partial_x v = 0,
\end{equation}
so $v$ is $x$-independent.  

The converse is also true: given any choice of $x$-independent volume 3-form $v$, a set of vector fields satisfying \eqref{div-free}, \eqref{nahm} can be used to assemble a hyperk\"ahler metric on the manifold $\mathbb{R} \times \cS$ via the formulas
\begin{equation}
\label{AJSresult}
h(V_\mu, V_\nu) = v(V_1, V_2, V_3) \, \delta_{\mu\nu}, \qquad X^i = \dd x \wedge h(V_i) + \ins_{V_i} v,
\end{equation}
where $\mu, \nu \in \{0,1,2,3\}$ and $V_0 \equiv \partial_x$ (in a coordinate chart $(x,y^i)$ where $y^i$ are coordinates on $\cS$). This enables one to construct the hyperk\"ahler manifold from data on $\cS$, at least in a neighbourhood of $\cS$. The method is the following. Let $v$ be a volume form on $\cS$. Let $t_i$ be three linearly-independent vector fields on $\cS$ that are divergence-free w.r.t. $v$. Now define $V_i$ to be vector fields satisfying the evolution equation \eqref{nahm} with initial conditions
\be
 V_i|_{x=0} = t_i
\ee
Standard theorems guarantee existence and uniqueness of a solution for $x \in (-\epsilon,\epsilon)$ for small enough $\epsilon>0$. The Nahm equations guarantee that the divergence-free condition \eqref{div-free} is preserved by the evolution. We now define $\cM$ to be the manifold $(-\epsilon,\epsilon) \times \cS$. In a coordinate chart $(x,y^i)$, where $y^i$ are coordinates on $\cS$, we define $V_0 = \partial/\partial x$. We extend $v$ onto $\cM$ by Lie transport w.r.t. $V_0$. The metric and hyper-k\"ahler 2-forms are then given by \eqref{AJSresult}.

Let us count the number of free functions in this data:  First fix a coordinate chart $y^i$ on $\cS$. Then each $t_i$ has $3$ free components but is subject to the condition that it preserves $v$. So each $t_i$ is equivalent to 2 free functions on $\cS$. Hence there are a total of $6$ free functions in the $3$ vector fields $t_i$.  However, we have gauge freedom associated to the freedom to choose the coordinates on $\cS$, which amounts to $3$ free function. Specifying $v$ appears to involve another free function but from \eqref{vdef} it can be seen that this is equivalent to the freedom to specify the normal derivative of the harmonic coordinate $x$ on $\cS$, i.e., a gauge degree of freedom. Finally, we could choose different locations for $\cS$ within the {\it same} hyperk\"ahler space; this gauge freedom to specify the location of $\cS$ is equivalent to another free function on $\cS$. So overall the number of non-gauge free functions on $\cS$ is $6-3-1=2$. This is equivalent to one `degree of freedom' in 4d, exactly as one would expect for a hyperk\"ahler space since such spaces are half-flat.

\section{Behavior of $f$ near $\cS$}
\label{app:general}

An important result needed in the arguments of \sref{sec:evanescent} is that $f$ should have a {\it first order} zero on an evanescent ergosurface.  The proof is in two steps:  first, demand that the Maxwell 2-form $F$ is smooth at $x=0$; second, demand that the Maxwell equation (i.e. the $f$ equation \eqref{f eqn}) is satisfied.

Begin with the expression for the Maxwell 2-form from \eqref{F expansion}, and put in $f = x^p \hat f$, where $\hat f$ is smooth and nonzero at $x \to 0$. We will later attempt to determine which values of $p$ are consistent with the Maxwell equation \eqref{f eqn}.  The Maxwell 2-form becomes
\begin{align}
2 \sqrt3 \, F &= - 3 \, \dd t \wedge \dd f \notag \\
\begin{split}
& + \dd x \wedge \nu \Big[ -p x^{-p-1} \hat f^{-1} - x^{-p} \hat f^{-2} \partial_x \hat f + 2 x^{-p} \hat f^{-1} (\dd \nu)_{x\nu} \\
& \qquad \qquad \qquad + x^{-2p} \hat f^{-2} (1+Qf^2)^{1/2} N (\dd \nu)_{12} \Big]
\end{split} \label{3r comp} \\
& + \rho^a \wedge \nu \Big[ - x^{-p} \hat f^{-2} e_a(\hat f) + 2 x^{-p} \hat f^{-1} (\dd \nu)_{a\nu} - x^{-2p} \hat f^{-2} (1+Qf^2)^{1/2} N \varepsilon^{ab} (\dd \nu)_{xb} \Big] \label{a3 comp} \\
&  + \dd x \wedge \rho^a \Big[ 2 x^{-p} \hat f^{-1} (\dd \nu)_{xa} + (1+Qf^2)^{-1/2} N \varepsilon^{ab} \Big( -2 \hat f^{-1} e_b(\hat f) + (\dd \nu)_{b\nu} \Big) \Big]. \notag \\
& + \rho^1 \wedge \rho^2 \Big[ 2 x^{-p} \hat f^{-1} (\dd \nu)_{12} - (1+Qf^2)^{-1/2} N^{-1} \Big( 2 p x^{-1} + 2 \hat f^{-1} \partial_x \hat f - (\dd\nu)_{x\nu} \Big) \Big] \notag
\end{align}
Since $p$ is variable, we will not try to work out \emph{all} the regularity conditions.  However, it is simple to work out the lowest-order conditions by cancelling only the most singular part.  From \eqref{3r comp} and \eqref{a3 comp}, we obtain
\begin{equation}
\label{P3 reg cond}
(\dd \nu)_{12} = p x^{p-1} \hat f + \cO(x^p), \qquad \text{and} \qquad (\dd \nu)_{xa} = x^p \varepsilon^{ab} \Big( e_b(\hat f) - 2 \hat f (\dd \nu)_{b\nu} \Big) + \cO(x^{p+1}).
\end{equation}
It will turn out that we will only require the first expression in \eqref{P3 reg cond}.

Now turn to the $f$ equation \eqref{f eqn} and apply the regularity condition \eqref{P3 reg cond}.  First we expand $\dd \hodge_4 \dd f^{-1}$ to remove the negative power of $f$ from under the derivatives, resulting in
\begin{equation} \label{f eqn rearranged}
\dd \hodge_4 \dd f = - \frac49 f^2 \, G^+ \wedge G^+ + 2 f^{-1} \, \dd f \wedge \hodge_4 \dd f.
\end{equation}
The left-hand side of \eqref{f eqn rearranged} is given by
\begin{equation} \label{LHS f eqn}
\begin{split}
\dd \hodge_4 \dd f &= \rho^1 \wedge \rho^2 \wedge \nu \wedge \dd x \, \times \\
& \qquad \times \Big \{ \partial_x \Big[ (1+Qf^2)^{1/2} N^{-1} \partial_x f \Big] - (1+Qf^2)^{1/2} N^{-1} \partial_x f \Big( P^a_{ax} - (\dd \nu)_{x\nu} \Big) \\
& \qquad \qquad + e_a \Big[ (1+Qf^2)^{1/2} N e_a(f) \Big] - (1+Qf^2)^{1/2} N e_a(f) \Big( P^b_{ba} - (\dd \nu)_{a\nu} \Big) \\
& \qquad \qquad + e_3 \Big[ f^2 (1+Qf^2)^{-1/2} N e_3(f) \Big] - f^2 (1+Qf^2)^{-1/2} N e_3(f) \Big( P^a_{a\nu} \Big) \Big \},
\end{split}
\end{equation}
where the quantities $P^a_{ij}$ are regular at $x \to 0$ and defined by
\begin{equation}
\dd \rho^a = P^a_{12} \, \rho^1 \wedge \rho^2 + P^a_{b\nu} \, \rho^b \wedge \nu + P^a_{bx} \, \rho^b \wedge \dd x + P^a_{x \nu} \, \dd x \wedge \nu.
\end{equation}
On the right-hand side of \eqref{f eqn rearranged}, we have\footnote{The notation $[\cdots]^{\underset{(a)}{2}}$ means that the expression is squared and summed over the index $a$; i.e. $[\cdots]_a [\cdots]^a$.}
\begin{equation}
\begin{split}
- \frac49 f^2 \, G^+ &\wedge G^+ + 2 f^{-1} \, \dd f \wedge \hodge_4 \dd f = \rho^1 \wedge \rho^2 \wedge \nu \wedge \dd x \, \times \\
& \qquad \times \Big \{ - \frac89 f^{-1} (1+Qf^2)^{1/2} N \Big[ \frac12 (\dd \nu)_{12} + (1+Qf^2)^{-1/2} N^{-1} \Big( \partial_x f - \frac12 f (\dd \nu)_{x\nu} \Big) \Big]^2 \\
& \qquad \qquad - \frac89 f^{-1} (1+Qf^2)^{1/2} N^{-1} \Big[ \frac12 (\dd \nu)_{xa} + (1+Qf^2)^{-1/2} N \varepsilon^{ab} \Big( e_b(f) - \frac12 f (\dd \nu)_{b\nu} \Big) \Big]^{\underset{(a)}{2}} \\
& \qquad \qquad + 2 f^{-1} (1+Qf^2)^{1/2} N^{-1} (\partial_r f)^2 + 2 f^{-1} (1+Qf^2)^{1/2} N e_a(f) e_a(f) \\
& \qquad \qquad + 2 f (1+Qf^2)^{-1/2} N \Big( e_3(f) \Big)^2 \Big \}.
\end{split}
\end{equation}
Putting $f = x^p \hat f$ and \eqref{P3 reg cond} into the above, and keeping only the lowest order terms, these expressions vastly simplify.  The $f$ equation \eqref{f eqn rearranged} becomes
\begin{equation}
p(p-1) x^{p-2} \hat f = - 2 p^2 x^{p-2} \hat f + 2 p^2 x^{p-2} \hat f + \cO(x^{p-1}) \quad = 0 + \cO(x^{p-1}).
\end{equation}
Therefore, if we demand that the Maxwell field $F$ is smooth and the $f$ equation is satisfied, to just \emph{one} order each, then we immediately conclude $p=1$, and thus $f$ must have a first order zero at $x=0$. 

\vskip 2cm
\bibliographystyle{utphys}
\bibliography{ambi_hk_draft}

\end{document}